\documentclass[
superscriptaddress,
amsmath,amssymb,amsfonts,aps,
longbibliography,
prb,
twocolumn,
]{revtex4-2}
\usepackage[T1]{fontenc}
\usepackage[utf8]{inputenc}
\usepackage{amstext, amsthm}
\usepackage{mathtools}
\usepackage{wasysym} 
\usepackage{scalerel}
\usepackage{tikz}
\usepackage{tikz-network}
\usetikzlibrary{patterns,decorations.pathreplacing}

\usepackage{latexsym,epsfig,graphicx}
\usepackage{dcolumn}

\usepackage{comment}
\usepackage{color}
\usepackage{bm}
\usepackage{mathrsfs}
\usepackage{tensor}
\usepackage{xspace, xcolor}
\usepackage{epstopdf}
\usepackage{tabularx}
\usepackage{lipsum} 
\usepackage[colorlinks=true, pdfstartview=FitV, linkcolor=red, citecolor=blue, urlcolor=blue]{hyperref}
\usepackage[normalem]{ulem}
\usepackage{natbib}
\usepackage[version=4]{mhchem}
\usepackage{braket}
\usepackage{url}
\usepackage{subfiles}

\definecolor{myred}{RGB}{232,102,102}
\definecolor{myblue}{RGB}{187,187,255}
\definecolor{myorange}{RGB}{255,165,0}
\definecolor{mygrey}{RGB}{105,105,105}
\definecolor{OliveGreen}{RGB}{85,107,47}
\definecolor{NavyBlue}{RGB}{0,0,128}
\definecolor{mygreen}{RGB}{34,139,34}
\definecolor{myY}{RGB}{220,255,203}
\definecolor{myYO}{RGB}{255, 220, 151}

\newcommand{\tr}{\mathrm{Tr}}
\newcommand{\lp}{\left( }
\newcommand{\rp}{\right) }

\makeatother

\begin{document}
	\title{Entanglement transition and suppression of critical phase of thermofield double state in monitored quantum circuit with unitary $R$ matrix gates}
	\author{Shi-Kang Sun}
	\email{sunsk@iphy.ac.cn}
	\affiliation{Beijing National Laboratory for Condensed Matter Physics, Institute of Physics, Chinese Academy of Sciences, Beijing 100190, China}
	\affiliation{School of Physical Sciences, University of Chinese Academy of Sciences, Beijing 100049, China}
	\author{Shu Chen}
	\email{schen@iphy.ac.cn }
	\affiliation{Beijing National Laboratory for Condensed Matter Physics, Institute of Physics, Chinese Academy of Sciences, Beijing 100190, China}
	\affiliation{School of Physical Sciences, University of Chinese Academy of Sciences, Beijing 100049, China}
	\date{\today}

	\begin{abstract}
		We study quantum circuits with gates composed randomly of identity operators, projectors, or a kind of $R$ matrices which satisfy the Yang-Baxter equation and are unitary and dual-unitary. This enables us to translate the quantum circuit into a topological object with distinguished overcrossings and undercrossings. The circuit corresponds to a classical loop model and is post-selection free when an overcrossing and an undercrossing coincide. The entanglement entropy between the final state and initial state is given by the spanning number of the classical model, and they share the same phase diagram. Whenever an overcrossing and undercrossing differ, the circuit extends beyond the classical model. Considering a specific case with $R$ matrices randomly replaced by SWAP gates, we demonstrate that the topological effect originating from worldline braiding dominates, and only the area-law phase remains in the thermodynamic limit, regardless of how small the replacement probability is. We also find evidence of an altered phase diagram for non-Clifford cases.
	\end{abstract}
	
	\maketitle
	
	\textit{Introduction}---
	Quantum circuits provide a natural platform to study discrete quantum evolution and are widely used in quantum computation, where time evolution is implemented as a sequence of quantum gates, and entanglement is a fundamental resource \citep{RevModPhys.81.865, RevModPhys.91.025001, PhysRevLett.102.190501}. One of the most intriguing characteristics of the quantum world is measurements, which also plays an important role in the circuit. Unitary circuits with measurements are often referred to as hybrid or monitored circuits. Measurements are non-unitary and give rise to novel nonequilibrium phenomena, for example, measurement-induced entanglement transitions \citep{PhysRevX.9.031009, PhysRevB.100.134306, RandomQuantumCircuits}, wherein the competition between unitary dynamics and projective measurements leads to fundamentally distinct entanglement scaling regimes.

	In recent years, there have been interesting connections between loop models and monitored circuits of Majorana and free fermion models \citep{PhysRevResearch.2.023288, PhysRevX.13.041028, PhysRevB.107.064303,PhysRevX.13.041045,PhysRevLett.133.070401,klocke2024entanglementdynamicsmonitoredkitaev}. The entanglement transitions therein are related to transitions of classical statistical models. In this work, we attempt to connect the realms of completely packed loop with crossings (CPLC) \citep{PhysRevB.87.184204, PhysRevLett.90.090601} with a spin-1/2 circuit, and take a step further to explore the effect of distinguishing different crossings, which is not encoded in the classical model and may contain novel phenomena resulting from topological effects.
	
	\begin{figure}[tb]
		\centering
		\includegraphics[width=0.7\linewidth]{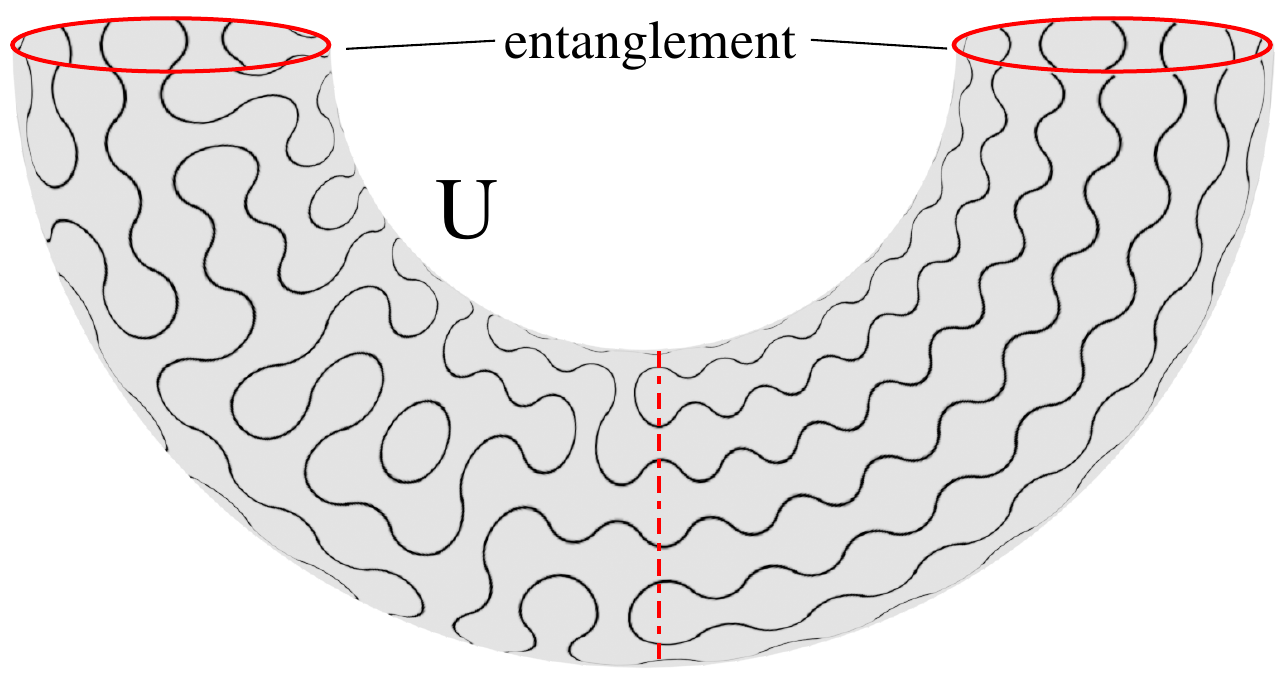}
		\caption{Entanglement protocol. $U$ is a quantum circuit. The time direction  goes upwards. The part on the right-hand side depicts an identity evolution of the right part of a thermofield double state $\ket{\Psi}$ (\ref{eq:tfds}). $U$ acts on the left part of $\ket{\Psi}$. We are interested in the entanglement between two parts, indicated by red circles at top.}
		\label{fig:3d1}
	\end{figure}
	
	Loop models are common in statistical physics. For example, they appear in the computation of the partition function of Ising model. Previous literature has studied the loop models with crossings in detail theoretically, relating them to supersymmetric spin chain, integrable models, and the $O(n)$ $\sigma$ model \citep{PhysRevLett.90.090601, READ2001409, PhysRevLett.81.504}, and numerically by Monte Carlo simulations \citep{Kager_2006, Ikhlef_2007}. Quantum loops are also used to model non-Abelian anyons \citep{FENDLEY20083113} and topological phases \citep{PhysRevB.71.045110}. The building blocks of loop models are connecting configurations of "strands" at a lattice site. A similar scenario also appears in integrable vertex systems \citep{baxter2016exactly}. Therefore, a natural way to map a loop model to a quantum circuit is to use unitary $R$ matrices \citep{Dye2003} from integrable systems as gates of a circuit \citep{PhysRevLett.122.150605, PhysRevB.103.115132}. $R$ matrices are solutions to the Yang-Baxter equation and also serve as generators of the braid group. They describe braiding of worldlines of particles, which is considered as a fault-tolerant method for topological quantum computation with anyons \citep{RevModPhys.80.1083}. By allowing unitary $R$ matrix gates, which are thus dual-unitary \citep{PhysRevLett.123.210601} and include standard SWAP gates as a special case, our circuit is non-orientable \citep{PhysRevX.13.041028} and can be always translated to a topological object, which has not been covered in previous studies. 
	We also generalize gates and entropy behaviors of \citep{PhysRevLett.126.060501} by topology, where a circuit with SWAP and identity gates is mapped to a loop model, and the pure phase is unstable for infinitesimal perturbations of SWAP gates randomly replaced by iSWAP gates.

	\textit{Circuit setup and entanglement}---
	In this work we are interested in the entanglement entropy between the initial and final state of a one-dimensional periodic spin 1/2 system, under a monitored quantum circuit with unitary $R$ matrix gates. This differs from usual entanglement entropy defined by a spatial cut of a quantum state, and is closer to what is known as pseudo entropy \citep{PhysRevD.103.026005} or temporal entanglement \citep{PhysRevA.91.032306, PhysRevX.11.021040, PhysRevX.13.041008, PhysRevResearch.6.043077}. It will be more convenient to explore the global (topological) property of the circuit when choosing this definition. The circuit $U$ used in this work generally relies on forced-measurements, but it becomes post-selection free and thus gets rid of this restriction once $U$ is Clifford as explained later. $U$ is a matrix product of unitary gates $R$, projectors $P$ and identity operators $I$ according to the brickwall pattern, $U = \prod_{t}\prod_{i=0,1}\otimes_{j}X^{t}_{2j-1+i, 2j+i}$ where $X^t$ is randomly chosen from $R$, $P$ and $I$ at time slice $t$, and the lower index indicates sites that $X^t$ acts on. The probability for unitary gates $R$ is $p$. At odd time layers the probability for measurement gates $P$ is $(1-p)q$ and $(1-p)(1-q)$ for identity gates $I$. $P$ and $I$ swap their probability on even time layers. This is the staggered probability setting used in \citep{PhysRevB.87.184204} as it respects the $P\leftrightarrow I$ symmetry, which is crucial for the transition therein. 
We can characterize the entanglement by "bending" the initial time boundary to be on the same time-plane as the final time boundary, viewing it as a replica of the original state. An illustration of this protocol is shown in Fig.~\ref{fig:3d1}.
	In this picture, the circuit $U$ acts on one side of an infinite-temperature thermofield double state (TFDS)
	\begin{equation}\label{eq:tfds}
			\ket{\Psi} = \frac{1}{\sqrt{2^L}} \otimes_{i=1}^L \lp \ket{\uparrow}_{i}\ket{\uparrow}_{i+L} + \ket{\downarrow}_{i}\ket{\downarrow}_{i+L}\rp 
			 \;\; .
	\end{equation}
	$\ket{\Psi}$ can also be written as $\frac{1}{\sqrt{2^L}} \sum_{x} \ket{x}_1\otimes \ket{x}_2$, where $L$ is the system size of either part and $\ket{x}$ is a computational basis, which is a product state of $\sigma_z$ eigenstates. We use subscript 1 and 2 to distinguish two parts, $(1, 2, \dots, L)$ and $(L+1, L+2, \dots, 2L)$, respectively. The state $\ket{\Psi}$ can be referred to as a pairing configuration, since every spin at site $i$ is paired with the spin at site $i+L$. 
	The entanglement of a single trajectory is defined as the von Neumann entropy $S=-\sum_i c_i^2 \log_2(c_i^2)$ of the time-evolved TFDS $U\otimes I \ket{\Psi} =\sum_{i} c_i \ket{\psi_i}_1 \otimes \ket{\phi_i}_2$ \citep{appendix_note}.
	
	\textit{Building block of the circuit}---
	To build a diagrammatic representation of the circuit, we need to find proper representations for all gates. First, we make $R$ matrix gates generators of a braid group, including standard swap gates in $\sigma_z$ basis as a special case.
	\begin{equation}
		R :=
		\raisebox{0.2em}{\begin{tikzpicture}[baseline=(current  bounding  box.center), scale=0.3]
				\draw[thick] (-1,-1) -- (1,1);
				\draw[thick] (-1, 1) -- (-0.2,0.2);
				\draw[thick](0.2, -0.2) -- (1,-1);
		\end{tikzpicture}}\; , \quad 
	R^{\dagger} :=
		\raisebox{0.2em}{\begin{tikzpicture}[baseline=(current  bounding  box.center), scale=0.3]
				\draw[thick] (-1,1) -- (1,-1);
				\draw[thick] (-1, -1) -- (-0.2,-0.2);
				\draw[thick](0.2, 0.2) -- (1,1);
		\end{tikzpicture}}
	\; , \quad
	\text{SWAP} := \raisebox{0.2em}{\begin{tikzpicture}[baseline=(current  bounding  box.center), scale=0.325]
			\draw[thick] (-1,-1) -- (1,1);
			\draw[thick] (-1, 1) -- (1,-1);
			\draw[thick, fill=myred] (0,0) circle [radius=0.2];
	\end{tikzpicture}} \;\; . 
	\end{equation}
	The SWAP gate is depicted as a crossing with a dot in the middle,  as it does not distinguish an over- or undercrossing.
	We define the Bell state in $\sigma_z$ basis,
	\begin{equation}\label{eq:bell_state}
		\ket{\uparrow\uparrow+\downarrow\downarrow}_{a,b}:=
		\raisebox{0.2em}{\begin{tikzpicture}[baseline=(current  bounding  box.center), scale=0.3]
				\Text[x=-1,y=-0.6]{\small $a$}
				\Text[x=1,y=-0.45]{\small $b$}
				\draw[thick] (-1,-1) arc [start angle= -180, end angle=0, radius=1];
		\end{tikzpicture}} \; ,  \quad
		\bra{\uparrow\uparrow+\downarrow\downarrow}_{a,b}:=
		\raisebox{0.2em}{\begin{tikzpicture}[baseline=(current  bounding  box.center), scale=0.3]
				\Text[x=-1,y=-1.1]{\small $a$}
				\Text[x=1,y=-1]{\small $b$}
				\draw[thick] (1,-0.6) arc [start angle= 0, end angle=180, radius=1];
		\end{tikzpicture}} \; , 
	\end{equation}
	so the (unnormalized) projector and identity operator are
	\begin{equation}\label{eq:graphical P and I}
		P' := \raisebox{0.2em}{\begin{tikzpicture}[baseline=(current  bounding  box.center), scale=0.6]
				\draw[thick] (-1,-1) arc [start angle= -135, end angle=-45, radius=0.707];
				\draw[thick] (-1,-2) arc [start angle= 135, end angle=45, radius=0.707];
		\end{tikzpicture}} = \ket{\uparrow\uparrow+\downarrow\downarrow}\bra{\uparrow\uparrow+\downarrow\downarrow}\; , \quad
		I := \raisebox{0.2em}{\begin{tikzpicture}[baseline=(current  bounding  box.center), scale=0.6]
				\draw[thick] (0,0) arc [start angle= -45, end angle=45, radius=0.707];
				\draw[thick] (1,0) arc [start angle= -135, end angle=-225, radius=0.707];
		\end{tikzpicture}}\; .
	\end{equation}
 $P' \equiv 2 P$ is not a projector since ${P'}^{2}=2 P'$. We interpret the lines in previous diagrammatic representations are worldlines of spins.
 
		In the literature, Kauffman's bracket \citep{10.1142/4256} uses the so-called skein relation to decompose a $R$ matrix as $R = AI+A^{-1}P'$, where $A$ is a free real number. However, when $R$ is unitary and local dimension $d=2$, $A$ is limited to $\pm i$. Instead, we introduce a family of unitary $R$ matrices that can be decomposed into three pieces with a real parameter $c$,
	\begin{equation}\label{eq:R_decom}
				R(c) = \frac{1}{\sqrt{1+c^2}} \lp i* I - i * P' + c * \text{SWAP}  \rp \;\;.
	\end{equation}The coefficients in (\ref{eq:R_decom}) are determined such that $R$ matrices are unitary two-qubit gates and are consistent with topological Reidemeister moves. Thus they are unitary, dual-unitary and satisfy the Yang-Baxter equation $R_{12}R_{23}R_{12}=R_{23}R_{12}R_{23}$, where subscripts indicate the sites the $R$ acting on. An explicit matrix representation of $R$ that we use in the circuit is
	\begin{equation}\label{eq:R_matrix}
		R(c)=\frac{1}{\sqrt{1+c^2}} \begin{pmatrix}
			c & 0 & 0 & -i \\
			0 & i & c & 0 \\
			0 & c & i & 0 \\
			-i & 0 & 0 & c
		\end{pmatrix} \;\; .
	\end{equation}
	Interestingly, under an unitary rotation $V$,  $V=e^{i \frac{\pi}{4}X_1}e^{i \frac{\pi}{4}X_2}$, the $R(c)$ gate reduces to $e^{-\frac{i \pi}{4}H}$ where $H$ is generally a two-site $XXZ$ hamiltonian, $H=X_1X_2+Y_1Y_2-\lp \frac{4}{\pi}\text{arccot} (c) -1 \rp Z_1Z_2-I_1I_2$. The $X$, $Y$ and $Z$ are Pauli operators. 
	The advantage for taking these $R$ gates reflects in the topological invariance of worldlines, as they can be transformed arbitrarily as long as their braiding topology is unchanged. We list some rules for the topological invariant:
\begin{equation}\label{eq:kpkm}
	\raisebox{0.2em}{\begin{tikzpicture}[baseline=(current  bounding  box.center), scale=0.25]
			\draw[thick] (0,0) circle [radius=1];
	\end{tikzpicture}} =n=2 \; , \quad
	\begin{tikzpicture}[baseline=(current  bounding  box.center), scale=0.25]
		\draw[thick] (-1,-1) -- (1,1);
		\draw[thick] (-1, 1) -- (-0.2,0.2);
		\draw[thick](0.2, -0.2) -- (1,-1);
		\draw[thick] (-1,-1) arc [start angle= -180, end angle=0, radius=1];
	\end{tikzpicture} = k_{+} \,\, \begin{tikzpicture}[baseline=(current  bounding  box.center), scale=0.25]
		\draw[thick] (-1,-1) -- (-1,1);
		\draw[thick] (1, 1) -- (1,-1);
		\draw[thick] (-1,-1) arc [start angle= -180, end angle=0, radius=1];
	\end{tikzpicture} \; , \quad
	\begin{tikzpicture}[baseline=(current  bounding  box.center), scale=0.25]
		\draw[thick] (-1,1) -- (1,-1);
		\draw[thick] (-1, -1) -- (-0.2,-0.2);
		\draw[thick](0.2, 0.2) -- (1,1);
		\draw[thick] (-1,-1) arc [start angle= -180, end angle=0, radius=1];
	\end{tikzpicture} = k_{-} \,\, \begin{tikzpicture}[baseline=(current  bounding  box.center), scale=0.25]
		\draw[thick] (-1,-1) -- (-1,1);
		\draw[thick] (1, 1) -- (1,-1);
		\draw[thick] (-1,-1) arc [start angle= -180, end angle=0, radius=1];
	\end{tikzpicture} \;\; ,
\end{equation}
where $n$ is loop value, $k_{+} = e^{-i \theta}$, $k_{-} = e^{i \theta}$, $\theta = \text{arccot}(c)$. They are related to the writhe of the corresponding knot. The rules regarding SWAP gates are:
\begin{equation}\label{eq:swap_gates}
	\begin{aligned}
		\begin{tikzpicture}[baseline=(current  bounding  box.center), scale=0.25]
			\draw[thick] (-1,-1) -- (1,1);
			\draw[thick] (-1, 1) -- (1,-1);
			\draw[thick] (-1,-1) arc [start angle= -180, end angle=0, radius=1];
			\draw[thick, fill=myred] (0,0) circle [radius= 0.2];
		\end{tikzpicture} = & \,\, \begin{tikzpicture}[baseline=(current  bounding  box.center), scale=0.25]
			\draw[thick] (-1,-1) -- (-1,1);
			\draw[thick] (1, 1) -- (1,-1);
			\draw[thick] (-1,-1) arc [start angle= -180, end angle=0, radius=1];
		\end{tikzpicture} \;, \quad
	\raisebox{0.2em}{\begin{tikzpicture}[baseline=(current  bounding  box.center), scale=0.3]
				\draw[thick] (-0.82,0) circle [radius=1];
				\draw[thick, dashed] (0.82,0) circle [radius=1];
				\draw[thick, fill=myred] (0,-0.64) circle [radius= 0.2];
		\end{tikzpicture}} = 
		\raisebox{0.2em}{\begin{tikzpicture}[baseline=(current  bounding  box.center), scale=0.3]
				\draw[thick] (-1,0) circle [radius=1];
				\draw[thick, dashed] (1,0) circle [radius=1];
				\draw[thick, fill=myred] (0,0) circle [radius= 0.2];
		\end{tikzpicture}} = g n^2 \; , \\
		&\raisebox{0.2em}{\begin{tikzpicture}[baseline=(current  bounding  box.center), scale=0.3]
				\draw[thick] (-0.75,0) circle [radius=1];
				\draw[thick, dashed] (0.75,0) circle [radius=1];
				\draw[thick, fill=myred] (0,0.66) circle [radius= 0.15];
				\draw[thick, fill=myred] (0,-0.66) circle [radius= 0.15];
		\end{tikzpicture}} =
		\raisebox{0.2em}{\begin{tikzpicture}[baseline=(current  bounding  box.center), scale=0.3]
				\draw[thick] (-1.25,0) circle [radius=1];
				\draw[thick, dashed] (1.25,0) circle [radius=1];
		\end{tikzpicture}} = n^2 \;\; ,
	\end{aligned}
\end{equation}
where $g=\frac{k_{+}+k_{-}}{2}=\frac{c}{\sqrt{1+c^2}}$ resulting from $R+R^{\dagger}$. The dashed lines are to indicate different connected components (loops). This rule states that if two distinct loops are concatenated by a dot, the dot contributes a multiplicative factor $g$. Given these rules, the topological invariant associated with every closed circuit configuration (periodic in space and time or inner product of two states) can be easily computed \citep{appendix_note}. For example, the second equation of (\ref{eq:swap_gates}) is proportional to the inner product $\braket{\psi |R^{\dagger}_{23} S_{23}|\psi}=g$, where $\ket{\psi}$ is $\frac{1}{2}\ket{\uparrow\uparrow+\downarrow\downarrow}_{12} \ket{\uparrow\uparrow+\downarrow\downarrow}_{34}$, $S$ is the SWAP gate and $c$ is the coefficient of $R(c)$.

During the circuit evolution, to make the mapping between graphical representation and quantum mechanics consistent, all measurements are forced measurements \textit{generally} so that the state after a measurement can still be represented by the worldline diagram. The state after a measurement is then $\ket{\psi'} = P\ket{\psi}$, and the new state looks like (\ref{eq:bell_state}) locally on the measured sites. We must also introduce an additional factor for the conversion between inner product of quantum states and the topological invariant of the worldline braiding of size $L$, 
\begin{equation}\label{eq:inner_prod}
		n^{\frac{L}{2}+\#(P)} \braket{\Phi | \Psi} = \tau(\Phi, \Psi) \leq n^{\frac{L}{2}+\#(P)} \;\; ,
\end{equation}where $\#(P)$ is the number of measurements in the circuit and $\tau$ is the topological invariant of the knot formed by concatenating two pieces of worldline configurations represented by $\bra{\Phi}$ and $\ket{\Psi}$.

	\textit{Topological effect on the entanglement}---
	\begin{figure}[tb]
		\centering
		\includegraphics[width=1.0\linewidth]{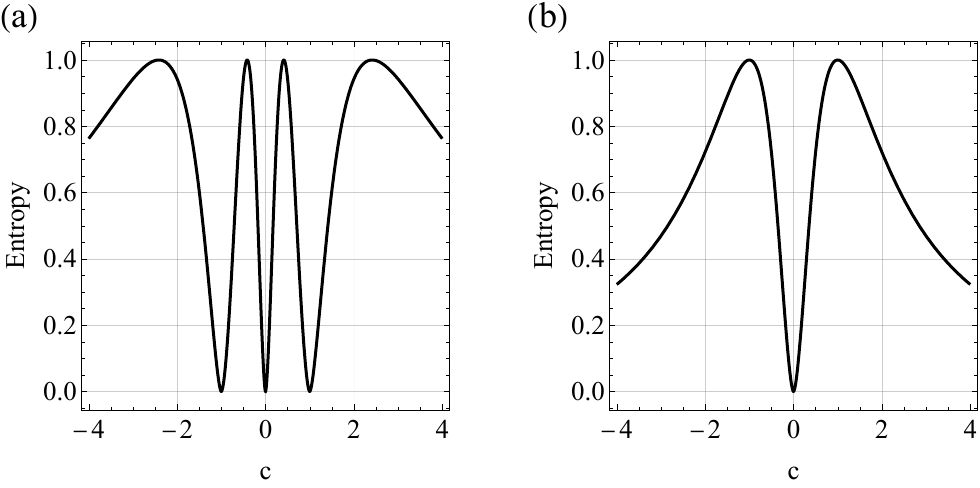}
		\caption{Entropy produced by a link of worldlines. (a) Link composed by two $R(c)$ matrix gates (\ref{eq:circuit-piece}a). The von Neumann entropy is zero at Clifford points and nonzero otherwise. (b) Link composed by one $R(c)$ matrix gate and one SWAP gate (\ref{eq:circuit-piece}b). The von Neumann entropy is not zero at $c=\pm 1$ as a result of different topology. }
		\label{fig:link_ent}
	\end{figure}The entanglement scaling behavior of the $R$ matrix circuit reveal its relation with CPLC and topological effect through very common structures in the circuit, worldline links. If two worldlines are tangled, for example (\ref{eq:circuit-piece}a), they cannot continuously pass through each other without changing the braiding topology, and therefore, entanglement.
	\begin{figure*}[tb]
		\centering
		\includegraphics[width=1.0\textwidth]{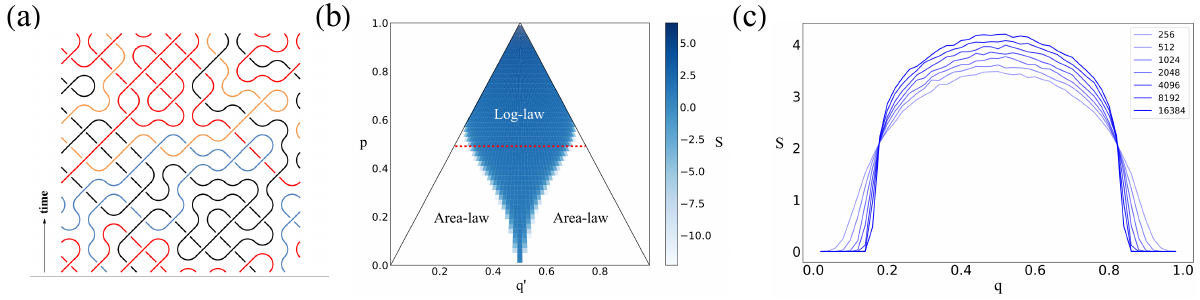}
		\caption{(a) A worldline configuration of loops with crossings. Red: worldlines that have both ends on the same boundary, but do not entangle with the other boundary. Black: worldlines that connect two boundaries and contribute to the spanning number. Blue and orange: worldlines that are entangled by a link. (b) Phase diagram of the entanglement entropy in log-scale for $R(c)$ set at Clifford points, computed for system size $L=2^{11}$, $t=L$ and are averaged from $10240$ samples. $p$ is the probability of $R$. $q'$ is defined by $(q'-1/2) = (q-1/2)(1-p)$ where $(1-p)q$ is the probability of $P$ on odd time layers. Blue part represents a critical region, where entanglement entropy grows logarithmically with system size $L$, and corresponds to the Goldstone phase of CPLC. White parts are area-law regions where entanglement entropy decays to 0 as system size $L$ grows and correspond to two short loop phases of CPLC. The red dashed line corresponds to data shown in (c). (c) Entanglement entropy of different system size when $p=0.5$.}
		\label{fig:outfig}
	\end{figure*}
We can check the von Neumann entropy associated with a link of worldlines by computing the entropy of a small circuit piece (\ref{eq:circuit-piece}a) that is common in any braiding,
	\begin{equation}\label{eq:circuit-piece}
		\text{(a)} \quad
		\begin{tikzpicture}[baseline=(current  bounding  box.center), scale=0.2]
			\draw[thick] (-1,-1) -- (1,1);
			\draw[thick] (-1, 1) -- (-0.2,0.2);
			\draw[thick](0.2, -0.2) -- (1,-1);
			
			\draw[thick] (3,-1) -- (5,1);
			\draw[thick] (3, 1) -- (3.8,0.2);
			\draw[thick](4.2, -0.2) -- (5,-1);
			
			\draw[thick] (3,1) arc [start angle= 45, end angle=135, radius=1.414];
			\draw[thick] (3,-1) arc [start angle= -45, end angle=-135, radius=1.414];
			
			\draw[thick] (3,-2.5) arc [start angle= 45, end angle=135, radius=1.414];
			\draw[thick] (3,2.5) arc [start angle= -45, end angle=-135, radius=1.414];
			
			\draw[thick] (-1,1) -- (-1,2.5);
			\draw[thick] (5,1) -- (5,2.5);
			
			\draw[thick] (-1,-1) -- (-1,-2.5);
			\draw[thick] (5,-1) -- (5,-2.5);
		\end{tikzpicture}\;\; ,	\quad		
			\text{(b)} \quad
			\begin{tikzpicture}[baseline=(current  bounding  box.center), scale=0.2]
				\draw[thick] (-1,-1) -- (1,1);
				\draw[thick] (-1, 1) -- (1,-1);
				\draw[thick, fill=myred] (0,0) circle [radius=0.2];
				
				\draw[thick] (3,-1) -- (5,1);
				\draw[thick] (3, 1) -- (3.8,0.2);
				\draw[thick](4.2, -0.2) -- (5,-1);
				
				\draw[thick] (3,1) arc [start angle= 45, end angle=135, radius=1.414];
				\draw[thick] (3,-1) arc [start angle= -45, end angle=-135, radius=1.414];
				
				\draw[thick] (3,-2.5) arc [start angle= 45, end angle=135, radius=1.414];
				\draw[thick] (3,2.5) arc [start angle= -45, end angle=-135, radius=1.414];
				
				\draw[thick] (-1,1) -- (-1,2.5);
				\draw[thick] (5,1) -- (5,2.5);
				
				\draw[thick] (-1,-1) -- (-1,-2.5);
				\draw[thick] (5,-1) -- (5,-2.5);
		\end{tikzpicture}
	\end{equation}
	and the result is shown in Fig.~\ref{fig:link_ent}(a). 
	The von Neumann entropy is zero only at Clifford points $c=0, \pm 1, \pm \infty$, where a crossing does not contribute to any entanglement and the circuit becomes a Clifford circuit \citep{gottesman1997stabilizer, PhysRevA.70.052328, PhysRevX.7.031016, PhysRevB.100.134306, Nielsen_Chuang_2010}, which means over- and undercrossings coincide. 
	One can use efficient stabilizer formalism in Clifford circuits. A pure state $\ket{\Psi}$ on $2L$ sites is determined by $2L$ mutually commuting Pauli strings of length $2L$. They generate a stabilizer group $\mathcal{G}$ and each string is a tensor product of Pauli operators. Thus we can focus on those Pauli strings to keep track of the circuit evolution \citep{appendix_note} because all gates map a Pauli string to another. The entanglement equals to $L - \log_2 |\mathcal{G}_1 |$, where $\log_2 |\mathcal{G}_1 |$ is the number of independent stabilizer group element supported only on part 1 \citep{PhysRevX.7.031016}. 
		At Clifford points, $R(c)$ gates behaves like "swap" of Pauli operators as long as $c$ is fixed. Measurements $P$ first project to $X_1 X_2$, and project to $Z_1 Z_2$ forthwith, and \textit{do not} need to be forced as long as we are measuring in $X_1 X_2$ and $Z_1 Z_2$ eigenstates, because different measuring result correspond to $\pm 1$ eigenvalues and have no effect on $\log_2 |\mathcal{G}_1 |$. Hence the circuit now is post-selection free.
		For initial state (\ref{eq:tfds}), $\log_2 |\mathcal{G}_1 | = 0$ since all nontrivial Pauli operators are separated by length $L$ and in different parts. So we can view every worldline as connecting a pair of nontrivial Pauli operators. If a measurement acts on a pair of worldlines that have endpoints within the same part, $\log_2 |\mathcal{G}_1 | $ increases by 2, and then the entanglement entropy decreases by 2. The entropy is thus given by the spanning number defined as the number of worldlines that connect initial and final time boundary that cannot be supported only on one side.  
	
For non-Clifford cases, we anticipate there will be more phases because the entanglement now depends also on link structures as a consequence of the topological effect of distinguishing over- and undercrossings. 
However, it is very hard to simulate non-Clifford cases because of the exponential growth of entanglement inside the circuit even via tensor networks. Fortunately, if one is allowed to replace a $R(c=1)$ matrix gate by a SWAP gate randomly, for example (\ref{eq:circuit-piece}b), which corresponds to the $\|c\| \rightarrow \infty$ case, we may reveal the topological effect and are still able to use the Clifford circuit method.

	\textit{CPLC behavior at Clifford points}---
	In the following, we restrict the circuit having a fixed aspect ratio 1, namely $t=L$, without loss of generality.
	When $R(c)$ is set at Clifford points, there is no difference between over- and undercrossings. Different worldlines can pass through each other freely, as long as their endpoints are fixed. The entanglement between the initial and final state depends solely on the spanning number. Since the probabilities of gates are independent from each other, the statistics of entanglement can be described by an $n=1$ CPLC, of which the partition function is
	$Z \equiv \sum_{c}  p^{n_R} [(1-p)q]^{n_m} [(1-p)(1-q)]^{n_I} = 1$, where $p$ is the probability of $R$ gates, $(1-p)q$ is the probability of measurement gates $P$, $n_m$, $n_I$, $n_R$ are the number of measurements, identity gates and $R$ gates, respectively. We sum up all configurations. An example of a configuration is shown in Fig.~\ref{fig:outfig}(a). The phase transition of CPLC directly translates to the entanglement transition of quantum circuit.
	
	If $p=0$, there are no crossings in the circuit, the dynamics can be described by the Temperley-Lieb algebra. The graphical representation of the circuit is a soup of bubbles that do not overlap. Since the entanglement entropy is given by the spanning number, we  anticipate a percolation transition in the spatial direction, and there will be no connected worldlines between the initial and final state. In this case, there is a zero entanglement phase produced by the percolation transition. By symmetry, the transition happens when projector probability $q=0.5$, so only area-law phases exist when $q\neq 0.5$.
	
Once unitary $R$ gates are added into the circuit, the dynamics is described by the Birman–Murakami–Wenzl (BMW) algebra \citep{appendix_note}. A new critical phase emerges at Clifford points $c=0, \pm 1, \pm \infty$, as shown in Fig.~\ref{fig:outfig}(b), and it matches the Goldstone phase of CPLC \citep{PhysRevB.87.184204, PhysRevLett.90.090601}, where entanglement grows logarithmically. We plot the entanglement entropy of different system sizes when $p=0.5$ in Fig.~\ref{fig:outfig}(c). Note that the point $p=1$ corresponds to a pure unitary circuit by $R$ matrix gates and thus retains maximum entanglement \citep{appendix_note}. The diagram boundary lines $q=0,1$ correspond also to area law phase, but entanglement increases as $p\rightarrow1$.

	\textit{Beyond CPLC: suppression of critical phase }\label{sec:topo_effect}---
	In the following, we will focus on the case where we replace  $R(c=1)$ matrix gates by SWAP gates randomly with probability $r$, depicted by (\ref{eq:circuit-piece}b). As shown in Fig.~\ref{fig:link_ent}(b), a link now contributes to some entanglement. Since a SWAP gate is a sum of $R$ and $R^\dagger$, the introduction of SWAP gates leads to a superposition of circuit configurations, which goes beyond the classical picture where the entanglement only rely on the spanning number of single configuration. Moreover, in terms of the topological invariant $\tau$, once a SWAP gate concatenates two distinct loops as in (\ref{eq:swap_gates}), it will bring an additional $g=\frac{1}{\sqrt{2}}<1$ factor into $\tau$. Varying $p$ only, $\tau$ decreases because there are more SWAP gates leading to more $g$ factors. 
		Note that the application of $U$ on $\ket{\Psi}$ leads to a naive superposition based on $R$ matrix decomposition and (\ref{eq:inner_prod}),
	\begin{equation}\label{eq:naive_decom}
		\ket{\Psi_f} \propto \sum_{\pi} \braket{\pi|\Psi_f}  \ket{\pi} \propto   \sum_{\pi} \tau\lp \pi, \Psi_f \rp  \ket{\pi} \;\; ,
	\end{equation}
	 since either $I$, $P'$ or the SWAP gate only changes a pairing configuration to another. We denote a general pairing configuration as $\ket{\pi} \propto \otimes_{i} \lp \ket{\uparrow}_{i}\ket{\uparrow}_{\pi(i)} + \ket{\downarrow}_{i}\ket{\downarrow}_{\pi(i)}\rp$ and the final state as $\ket{\Psi_f} \equiv U \otimes I \ket{\Psi} $. In other words, the final state is in the space spanned by different pairing configurations, although they are not orthogonal. The overlapping coefficients relate to the topological invariant $\tau$, and their norm are all the same for a stabilizer state \citep{sierant2022universal}. So if $\tau$ decreases, coefficients also decrease, which means $\ket{\Psi_f}$  must be supported on a larger subspace because it is normalized by definition. We verify this by numerically computing the participation entropy \citep{sierant2022universal} $S_{\text{par}}$ of $\ket{\Psi_f}$ of size $2L=2048$. $S_{\text{par}}$ is an integer ranging from 0 to $2L$ and characterizes the distribution of a stabilizer state over the computational basis: when $S_{\text{par}} = 0$, $\ket{\Psi_f}$ is fully localized and when $S_{\text{par}} = 2L$, $\ket{\Psi_f}$ is fully extended.
	 	Fixing $r=0.1$, $q=0.5$, $S_{\text{par}}/2L$ equals to 0.59, 0.68, 0.82, 0.92 for $p=0.1, 0.3, 0.6, 0.8$, respectively. 
 	
\begin{figure}[tb]
		\centering
		\includegraphics[width=1.0\linewidth]{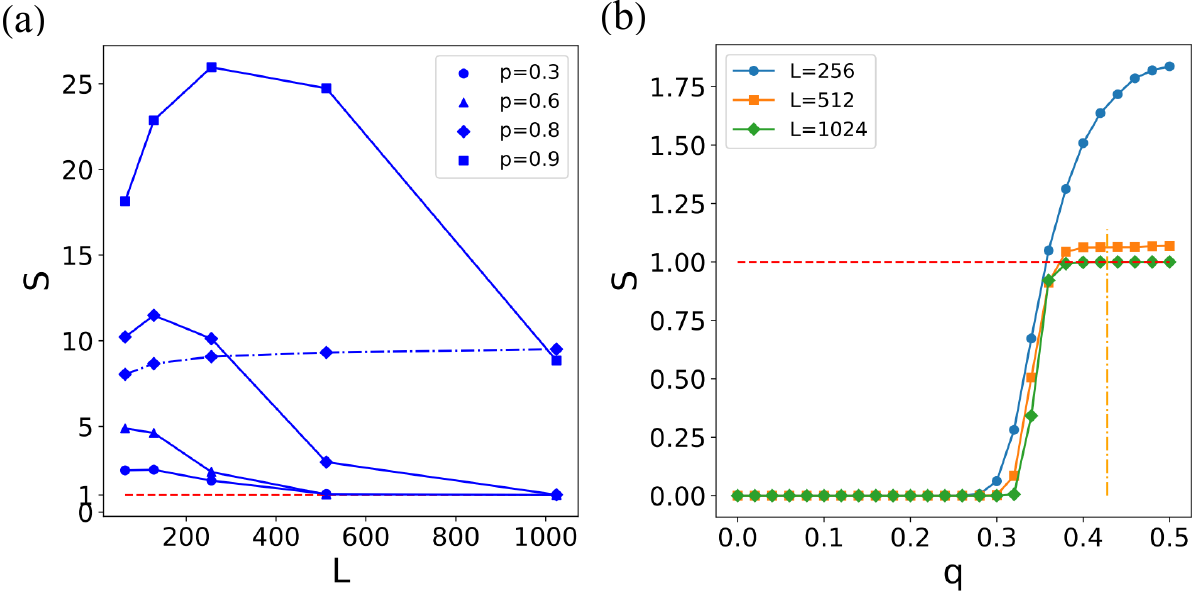}
		\caption{Red dashed line is a reference line for $S=1$. All data points are averaged from 16384 samples. (a) System size scaling of average entanglement entropy at time $t=L$ for $q=0.5$, $r=0.1$, which corresponds to the critical region of CPLC. Data shows convergence to $S=1$. The dash-dot line corresponds to $p=0.8$, $r=0$. (b) Averaged entanglement for $p=0.3$, $r=0.1$, $t=L=256, 512, 1024$.  Orange dash-dot line marks the transition point of CPLC.}
		\label{fig:entp0d5-lscaling}
	\end{figure}

 	Note that SWAP gates conserve the total parity $P^z = \prod_i^{2L} Z_i$ and $P^x= \prod_i^{2L} X_i$ of a state and projectors project the state into a locally paired state, then according to (\ref{eq:naive_decom}), the final state $\ket{\Psi_f}$ under our random circuit should be an even parity state as the initial state (a pairing configuration state) and has no other conserved quantities after random replacements of $R$ gates by SWAP gates. If parameters $p$, $q$ are set in critical region, the spanning number is not zero and both parts of $\ket{\Psi_f}$ are correlated, which means both parts can have odd or even parity, as long as the total parity is even. So $\ket{\Psi_f}$ has at least 1 unit of entanglement, since given conserved $P^z$ and $P^x$, the number of independent stabilizer group element supported only on part 1 or 2 satisfies $\log_2 |\mathcal{G}_1 | = \log_2 |\mathcal{G}_2 | \leq L-1$. 
	Numerical result shows the entanglement entropy $\langle S \rangle \rightarrow 1$ in the thermodynamic limit, as shown in Fig.~\ref{fig:entp0d5-lscaling}. For comparison, we show data for $r=0$ where entanglement keeps increasing. We \textit{conjecture} the final state approaches a sum of some odd and even parity states,  
	\begin{equation}
		\ket{\Psi_f} \rightarrow \lp \ket{\psi_\text{odd}}_1 \otimes \ket{\phi_\text{odd}}_2 + \ket{\psi^{'}_\text{even}}_1 \otimes \ket{\phi^{'}_\text{even}}_2\rp, 
\end{equation}
where $\ket{\psi_\text{even}}$ is a sum of some even parity states, namely $\ket{\psi_\text{even}} \propto \sum_{j \in (1,\dots 2^L)} a_j \ket{\prod^L_i Z_i = 1}_j $, $\| a_j \| = 1$.

	So after adding some random replacements of SWAP gates, the logarithmic increment of entanglement is suppressed. The reminiscent of the original critical phase is the final entanglement convergence to $1$ in the thermodynamic limit. However, some regions of original area law phase, that are near original transition lines, are altered by the replacements as shown in Fig.~\ref{fig:entp0d5-lscaling}(b). In those regions, instead of 0, they acquire some entanglement close to 1. Since in those regions the spanning number is 0 when $L \rightarrow \infty $ and contributes no entanglement, we \textit{conjecture} the entanglement behavior in those regions is related to some structures like chained rings, which could lead to a modified phase boundary in non-Clifford cases \citep{appendix_note}. In contrast, the remaining regions far away from original phase boundaries are unaltered, and the converged value of entanglement is 0 because those two boundaries are totally disconnected as we can draw a line to separate them without breaking any worldline. Thus, the final state can be written as $\ket{\Psi_f} \rightarrow \ket{\psi_\text{even}}\otimes \ket{\phi_\text{even}}$. 
	
	\textit{Conclusion}---
	We construct a 4-dimensional matrix representation of the braid group generator $R$. We study the averaged entanglement between the initial and final state under the circuit evolution made by these $R$ matrices and relevant gates. When the circuit is Clifford the circuit can be described by CPLC and entanglement shares the same phase diagram. If some $R$ gates are replaced by SWAP gates, the critical phase is suppressed by a topological effect, in contrast to previous result under general Gaussian unitary gates \citep{PhysRevB.107.064303}. The averaged entanglement also shows different behavior beyond CPLC for regions of area-law phases near phase boundaries. We conjecture they are evidences that the phase diagram will be enhanced by $R$ matrix gates outside Clifford points.
	
	\textit{Acknowledgments}---
		We thank K. Wang, C. G. Liang, R. Qi, and X. Feng for helpful discussions. Code and data can be found at \citep{data}. This work is supported by National Key Research and Development Program of China (Grant No. 2023YFA1406704)
and the NSFC under Grants  No.12474287  and No. T2121001.

	\bibliography{libbib}

\begin{thebibliography}{40}%
\makeatletter
\providecommand \@ifxundefined [1]{%
 \@ifx{#1\undefined}
}%
\providecommand \@ifnum [1]{%
 \ifnum #1\expandafter \@firstoftwo
 \else \expandafter \@secondoftwo
 \fi
}%
\providecommand \@ifx [1]{%
 \ifx #1\expandafter \@firstoftwo
 \else \expandafter \@secondoftwo
 \fi
}%
\providecommand \natexlab [1]{#1}%
\providecommand \enquote  [1]{``#1''}%
\providecommand \bibnamefont  [1]{#1}%
\providecommand \bibfnamefont [1]{#1}%
\providecommand \citenamefont [1]{#1}%
\providecommand \href@noop [0]{\@secondoftwo}%
\providecommand \href [0]{\begingroup \@sanitize@url \@href}%
\providecommand \@href[1]{\@@startlink{#1}\@@href}%
\providecommand \@@href[1]{\endgroup#1\@@endlink}%
\providecommand \@sanitize@url [0]{\catcode `\\12\catcode `\$12\catcode
  `\&12\catcode `\#12\catcode `\^12\catcode `\_12\catcode `\%12\relax}%
\providecommand \@@startlink[1]{}%
\providecommand \@@endlink[0]{}%
\providecommand \url  [0]{\begingroup\@sanitize@url \@url }%
\providecommand \@url [1]{\endgroup\@href {#1}{\urlprefix }}%
\providecommand \urlprefix  [0]{URL }%
\providecommand \Eprint [0]{\href }%
\providecommand \doibase [0]{https://doi.org/}%
\providecommand \selectlanguage [0]{\@gobble}%
\providecommand \bibinfo  [0]{\@secondoftwo}%
\providecommand \bibfield  [0]{\@secondoftwo}%
\providecommand \translation [1]{[#1]}%
\providecommand \BibitemOpen [0]{}%
\providecommand \bibitemStop [0]{}%
\providecommand \bibitemNoStop [0]{.\EOS\space}%
\providecommand \EOS [0]{\spacefactor3000\relax}%
\providecommand \BibitemShut  [1]{\csname bibitem#1\endcsname}%
\let\auto@bib@innerbib\@empty
\bibitem [{\citenamefont {Horodecki}\ \emph {et~al.}(2009)\citenamefont
  {Horodecki}, \citenamefont {Horodecki}, \citenamefont {Horodecki},\ and\
  \citenamefont {Horodecki}}]{RevModPhys.81.865}%
  \BibitemOpen
  \bibfield  {author} {\bibinfo {author} {\bibfnamefont {R.}~\bibnamefont
  {Horodecki}}, \bibinfo {author} {\bibfnamefont {P.}~\bibnamefont
  {Horodecki}}, \bibinfo {author} {\bibfnamefont {M.}~\bibnamefont
  {Horodecki}},\ and\ \bibinfo {author} {\bibfnamefont {K.}~\bibnamefont
  {Horodecki}},\ }\bibfield  {title} {\bibinfo {title} {Quantum entanglement},\
  }\href {https://doi.org/10.1103/RevModPhys.81.865} {\bibfield  {journal}
  {\bibinfo  {journal} {Rev. Mod. Phys.}\ }\textbf {\bibinfo {volume} {81}},\
  \bibinfo {pages} {865} (\bibinfo {year} {2009})}\BibitemShut {NoStop}%
\bibitem [{\citenamefont {Chitambar}\ and\ \citenamefont
  {Gour}(2019)}]{RevModPhys.91.025001}%
  \BibitemOpen
  \bibfield  {author} {\bibinfo {author} {\bibfnamefont {E.}~\bibnamefont
  {Chitambar}}\ and\ \bibinfo {author} {\bibfnamefont {G.}~\bibnamefont
  {Gour}},\ }\bibfield  {title} {\bibinfo {title} {Quantum resource theories},\
  }\href {https://doi.org/10.1103/RevModPhys.91.025001} {\bibfield  {journal}
  {\bibinfo  {journal} {Rev. Mod. Phys.}\ }\textbf {\bibinfo {volume} {91}},\
  \bibinfo {pages} {025001} (\bibinfo {year} {2019})}\BibitemShut {NoStop}%
\bibitem [{\citenamefont {Gross}\ \emph {et~al.}(2009)\citenamefont {Gross},
  \citenamefont {Flammia},\ and\ \citenamefont
  {Eisert}}]{PhysRevLett.102.190501}%
  \BibitemOpen
  \bibfield  {author} {\bibinfo {author} {\bibfnamefont {D.}~\bibnamefont
  {Gross}}, \bibinfo {author} {\bibfnamefont {S.~T.}\ \bibnamefont {Flammia}},\
  and\ \bibinfo {author} {\bibfnamefont {J.}~\bibnamefont {Eisert}},\
  }\bibfield  {title} {\bibinfo {title} {Most quantum states are too entangled
  to be useful as computational resources},\ }\href
  {https://doi.org/10.1103/PhysRevLett.102.190501} {\bibfield  {journal}
  {\bibinfo  {journal} {Phys. Rev. Lett.}\ }\textbf {\bibinfo {volume} {102}},\
  \bibinfo {pages} {190501} (\bibinfo {year} {2009})}\BibitemShut {NoStop}%
\bibitem [{\citenamefont {Skinner}\ \emph {et~al.}(2019)\citenamefont
  {Skinner}, \citenamefont {Ruhman},\ and\ \citenamefont
  {Nahum}}]{PhysRevX.9.031009}%
  \BibitemOpen
  \bibfield  {author} {\bibinfo {author} {\bibfnamefont {B.}~\bibnamefont
  {Skinner}}, \bibinfo {author} {\bibfnamefont {J.}~\bibnamefont {Ruhman}},\
  and\ \bibinfo {author} {\bibfnamefont {A.}~\bibnamefont {Nahum}},\ }\bibfield
   {title} {\bibinfo {title} {Measurement-induced phase transitions in the
  dynamics of entanglement},\ }\href
  {https://doi.org/10.1103/PhysRevX.9.031009} {\bibfield  {journal} {\bibinfo
  {journal} {Phys. Rev. X}\ }\textbf {\bibinfo {volume} {9}},\ \bibinfo {pages}
  {031009} (\bibinfo {year} {2019})}\BibitemShut {NoStop}%
\bibitem [{\citenamefont {Li}\ \emph {et~al.}(2019)\citenamefont {Li},
  \citenamefont {Chen},\ and\ \citenamefont {Fisher}}]{PhysRevB.100.134306}%
  \BibitemOpen
  \bibfield  {author} {\bibinfo {author} {\bibfnamefont {Y.}~\bibnamefont
  {Li}}, \bibinfo {author} {\bibfnamefont {X.}~\bibnamefont {Chen}},\ and\
  \bibinfo {author} {\bibfnamefont {M.~P.~A.}\ \bibnamefont {Fisher}},\
  }\bibfield  {title} {\bibinfo {title} {Measurement-driven entanglement
  transition in hybrid quantum circuits},\ }\href
  {https://doi.org/10.1103/PhysRevB.100.134306} {\bibfield  {journal} {\bibinfo
   {journal} {Phys. Rev. B}\ }\textbf {\bibinfo {volume} {100}},\ \bibinfo
  {pages} {134306} (\bibinfo {year} {2019})}\BibitemShut {NoStop}%
\bibitem [{\citenamefont {Fisher}\ \emph {et~al.}(2023)\citenamefont {Fisher},
  \citenamefont {Khemani}, \citenamefont {Nahum},\ and\ \citenamefont
  {Vijay}}]{RandomQuantumCircuits}%
  \BibitemOpen
  \bibfield  {author} {\bibinfo {author} {\bibfnamefont {M.~P.}\ \bibnamefont
  {Fisher}}, \bibinfo {author} {\bibfnamefont {V.}~\bibnamefont {Khemani}},
  \bibinfo {author} {\bibfnamefont {A.}~\bibnamefont {Nahum}},\ and\ \bibinfo
  {author} {\bibfnamefont {S.}~\bibnamefont {Vijay}},\ }\bibfield  {title}
  {\bibinfo {title} {Random quantum circuits},\ }\href
  {https://doi.org/https://doi.org/10.1146/annurev-conmatphys-031720-030658}
  {\bibfield  {journal} {\bibinfo  {journal} {Annual Review of Condensed Matter
  Physics}\ }\textbf {\bibinfo {volume} {14}},\ \bibinfo {pages} {335}
  (\bibinfo {year} {2023})}\BibitemShut {NoStop}%
\bibitem [{\citenamefont {Nahum}\ and\ \citenamefont
  {Skinner}(2020)}]{PhysRevResearch.2.023288}%
  \BibitemOpen
  \bibfield  {author} {\bibinfo {author} {\bibfnamefont {A.}~\bibnamefont
  {Nahum}}\ and\ \bibinfo {author} {\bibfnamefont {B.}~\bibnamefont
  {Skinner}},\ }\bibfield  {title} {\bibinfo {title} {Entanglement and dynamics
  of diffusion-annihilation processes with majorana defects},\ }\href
  {https://doi.org/10.1103/PhysRevResearch.2.023288} {\bibfield  {journal}
  {\bibinfo  {journal} {Phys. Rev. Res.}\ }\textbf {\bibinfo {volume} {2}},\
  \bibinfo {pages} {023288} (\bibinfo {year} {2020})}\BibitemShut {NoStop}%
\bibitem [{\citenamefont {Klocke}\ and\ \citenamefont
  {Buchhold}(2023)}]{PhysRevX.13.041028}%
  \BibitemOpen
  \bibfield  {author} {\bibinfo {author} {\bibfnamefont {K.}~\bibnamefont
  {Klocke}}\ and\ \bibinfo {author} {\bibfnamefont {M.}~\bibnamefont
  {Buchhold}},\ }\bibfield  {title} {\bibinfo {title} {Majorana loop models for
  measurement-only quantum circuits},\ }\href
  {https://doi.org/10.1103/PhysRevX.13.041028} {\bibfield  {journal} {\bibinfo
  {journal} {Phys. Rev. X}\ }\textbf {\bibinfo {volume} {13}},\ \bibinfo
  {pages} {041028} (\bibinfo {year} {2023})}\BibitemShut {NoStop}%
\bibitem [{\citenamefont {Merritt}\ and\ \citenamefont
  {Fidkowski}(2023)}]{PhysRevB.107.064303}%
  \BibitemOpen
  \bibfield  {author} {\bibinfo {author} {\bibfnamefont {J.}~\bibnamefont
  {Merritt}}\ and\ \bibinfo {author} {\bibfnamefont {L.}~\bibnamefont
  {Fidkowski}},\ }\bibfield  {title} {\bibinfo {title} {Entanglement
  transitions with free fermions},\ }\href
  {https://doi.org/10.1103/PhysRevB.107.064303} {\bibfield  {journal} {\bibinfo
   {journal} {Phys. Rev. B}\ }\textbf {\bibinfo {volume} {107}},\ \bibinfo
  {pages} {064303} (\bibinfo {year} {2023})}\BibitemShut {NoStop}%
\bibitem [{\citenamefont {Fava}\ \emph {et~al.}(2023)\citenamefont {Fava},
  \citenamefont {Piroli}, \citenamefont {Swann}, \citenamefont {Bernard},\ and\
  \citenamefont {Nahum}}]{PhysRevX.13.041045}%
  \BibitemOpen
  \bibfield  {author} {\bibinfo {author} {\bibfnamefont {M.}~\bibnamefont
  {Fava}}, \bibinfo {author} {\bibfnamefont {L.}~\bibnamefont {Piroli}},
  \bibinfo {author} {\bibfnamefont {T.}~\bibnamefont {Swann}}, \bibinfo
  {author} {\bibfnamefont {D.}~\bibnamefont {Bernard}},\ and\ \bibinfo {author}
  {\bibfnamefont {A.}~\bibnamefont {Nahum}},\ }\bibfield  {title} {\bibinfo
  {title} {Nonlinear sigma models for monitored dynamics of free fermions},\
  }\href {https://doi.org/10.1103/PhysRevX.13.041045} {\bibfield  {journal}
  {\bibinfo  {journal} {Phys. Rev. X}\ }\textbf {\bibinfo {volume} {13}},\
  \bibinfo {pages} {041045} (\bibinfo {year} {2023})}\BibitemShut {NoStop}%
\bibitem [{\citenamefont {Klocke}\ \emph
  {et~al.}(2024{\natexlab{a}})\citenamefont {Klocke}, \citenamefont {Moore},\
  and\ \citenamefont {Buchhold}}]{PhysRevLett.133.070401}%
  \BibitemOpen
  \bibfield  {author} {\bibinfo {author} {\bibfnamefont {K.}~\bibnamefont
  {Klocke}}, \bibinfo {author} {\bibfnamefont {J.~E.}\ \bibnamefont {Moore}},\
  and\ \bibinfo {author} {\bibfnamefont {M.}~\bibnamefont {Buchhold}},\
  }\bibfield  {title} {\bibinfo {title} {Power-law entanglement and hilbert
  space fragmentation in nonreciprocal quantum circuits},\ }\href
  {https://doi.org/10.1103/PhysRevLett.133.070401} {\bibfield  {journal}
  {\bibinfo  {journal} {Phys. Rev. Lett.}\ }\textbf {\bibinfo {volume} {133}},\
  \bibinfo {pages} {070401} (\bibinfo {year} {2024}{\natexlab{a}})}\BibitemShut
  {NoStop}%
\bibitem [{\citenamefont {Klocke}\ \emph
  {et~al.}(2024{\natexlab{b}})\citenamefont {Klocke}, \citenamefont {Simm},
  \citenamefont {Zhu}, \citenamefont {Trebst},\ and\ \citenamefont
  {Buchhold}}]{klocke2024entanglementdynamicsmonitoredkitaev}%
  \BibitemOpen
  \bibfield  {author} {\bibinfo {author} {\bibfnamefont {K.}~\bibnamefont
  {Klocke}}, \bibinfo {author} {\bibfnamefont {D.}~\bibnamefont {Simm}},
  \bibinfo {author} {\bibfnamefont {G.-Y.}\ \bibnamefont {Zhu}}, \bibinfo
  {author} {\bibfnamefont {S.}~\bibnamefont {Trebst}},\ and\ \bibinfo {author}
  {\bibfnamefont {M.}~\bibnamefont {Buchhold}},\ }\href
  {https://arxiv.org/abs/2409.02171} {\bibinfo {title} {Entanglement dynamics
  in monitored kitaev circuits: loop models, symmetry classification, and
  quantum lifshitz scaling}} (\bibinfo {year} {2024}{\natexlab{b}}),\ \Eprint
  {https://arxiv.org/abs/2409.02171} {arXiv:2409.02171} \BibitemShut {NoStop}%
\bibitem [{\citenamefont {Nahum}\ \emph {et~al.}(2013)\citenamefont {Nahum},
  \citenamefont {Serna}, \citenamefont {Somoza},\ and\ \citenamefont
  {Ortu\~no}}]{PhysRevB.87.184204}%
  \BibitemOpen
  \bibfield  {author} {\bibinfo {author} {\bibfnamefont {A.}~\bibnamefont
  {Nahum}}, \bibinfo {author} {\bibfnamefont {P.}~\bibnamefont {Serna}},
  \bibinfo {author} {\bibfnamefont {A.~M.}\ \bibnamefont {Somoza}},\ and\
  \bibinfo {author} {\bibfnamefont {M.}~\bibnamefont {Ortu\~no}},\ }\bibfield
  {title} {\bibinfo {title} {Loop models with crossings},\ }\href
  {https://doi.org/10.1103/PhysRevB.87.184204} {\bibfield  {journal} {\bibinfo
  {journal} {Phys. Rev. B}\ }\textbf {\bibinfo {volume} {87}},\ \bibinfo
  {pages} {184204} (\bibinfo {year} {2013})}\BibitemShut {NoStop}%
\bibitem [{\citenamefont {Jacobsen}\ \emph {et~al.}(2003)\citenamefont
  {Jacobsen}, \citenamefont {Read},\ and\ \citenamefont
  {Saleur}}]{PhysRevLett.90.090601}%
  \BibitemOpen
  \bibfield  {author} {\bibinfo {author} {\bibfnamefont {J.~L.}\ \bibnamefont
  {Jacobsen}}, \bibinfo {author} {\bibfnamefont {N.}~\bibnamefont {Read}},\
  and\ \bibinfo {author} {\bibfnamefont {H.}~\bibnamefont {Saleur}},\
  }\bibfield  {title} {\bibinfo {title} {Dense loops, supersymmetry, and
  goldstone phases in two dimensions},\ }\href
  {https://doi.org/10.1103/PhysRevLett.90.090601} {\bibfield  {journal}
  {\bibinfo  {journal} {Phys. Rev. Lett.}\ }\textbf {\bibinfo {volume} {90}},\
  \bibinfo {pages} {090601} (\bibinfo {year} {2003})}\BibitemShut {NoStop}%
\bibitem [{\citenamefont {Read}\ and\ \citenamefont
  {Saleur}(2001)}]{READ2001409}%
  \BibitemOpen
  \bibfield  {author} {\bibinfo {author} {\bibfnamefont {N.}~\bibnamefont
  {Read}}\ and\ \bibinfo {author} {\bibfnamefont {H.}~\bibnamefont {Saleur}},\
  }\bibfield  {title} {\bibinfo {title} {Exact spectra of conformal
  supersymmetric nonlinear sigma models in two dimensions},\ }\href
  {https://doi.org/https://doi.org/10.1016/S0550-3213(01)00395-9} {\bibfield
  {journal} {\bibinfo  {journal} {Nuclear Physics B}\ }\textbf {\bibinfo
  {volume} {613}},\ \bibinfo {pages} {409} (\bibinfo {year}
  {2001})}\BibitemShut {NoStop}%
\bibitem [{\citenamefont {Martins}\ \emph {et~al.}(1998)\citenamefont
  {Martins}, \citenamefont {Nienhuis},\ and\ \citenamefont
  {Rietman}}]{PhysRevLett.81.504}%
  \BibitemOpen
  \bibfield  {author} {\bibinfo {author} {\bibfnamefont {M.~J.}\ \bibnamefont
  {Martins}}, \bibinfo {author} {\bibfnamefont {B.}~\bibnamefont {Nienhuis}},\
  and\ \bibinfo {author} {\bibfnamefont {R.}~\bibnamefont {Rietman}},\
  }\bibfield  {title} {\bibinfo {title} {Intersecting loop model as a solvable
  super spin chain},\ }\href {https://doi.org/10.1103/PhysRevLett.81.504}
  {\bibfield  {journal} {\bibinfo  {journal} {Phys. Rev. Lett.}\ }\textbf
  {\bibinfo {volume} {81}},\ \bibinfo {pages} {504} (\bibinfo {year}
  {1998})}\BibitemShut {NoStop}%
\bibitem [{\citenamefont {Kager}\ and\ \citenamefont
  {Nienhuis}(2006)}]{Kager_2006}%
  \BibitemOpen
  \bibfield  {author} {\bibinfo {author} {\bibfnamefont {W.}~\bibnamefont
  {Kager}}\ and\ \bibinfo {author} {\bibfnamefont {B.}~\bibnamefont
  {Nienhuis}},\ }\bibfield  {title} {\bibinfo {title} {Monte carlo study of the
  hull distribution for the q = 1 brauer model},\ }\href
  {https://doi.org/10.1088/1742-5468/2006/08/P08004} {\bibfield  {journal}
  {\bibinfo  {journal} {Journal of Statistical Mechanics: Theory and
  Experiment}\ }\textbf {\bibinfo {volume} {2006}},\ \bibinfo {pages} {P08004}
  (\bibinfo {year} {2006})}\BibitemShut {NoStop}%
\bibitem [{\citenamefont {Ikhlef}\ \emph {et~al.}(2007)\citenamefont {Ikhlef},
  \citenamefont {Jacobsen},\ and\ \citenamefont {Saleur}}]{Ikhlef_2007}%
  \BibitemOpen
  \bibfield  {author} {\bibinfo {author} {\bibfnamefont {Y.}~\bibnamefont
  {Ikhlef}}, \bibinfo {author} {\bibfnamefont {J.}~\bibnamefont {Jacobsen}},\
  and\ \bibinfo {author} {\bibfnamefont {H.}~\bibnamefont {Saleur}},\
  }\bibfield  {title} {\bibinfo {title} {Non-intersection exponents of fully
  packed trails on the square lattice},\ }\href
  {https://doi.org/10.1088/1742-5468/2007/05/P05005} {\bibfield  {journal}
  {\bibinfo  {journal} {Journal of Statistical Mechanics: Theory and
  Experiment}\ }\textbf {\bibinfo {volume} {2007}},\ \bibinfo {pages} {P05005}
  (\bibinfo {year} {2007})}\BibitemShut {NoStop}%
\bibitem [{\citenamefont {Fendley}(2008)}]{FENDLEY20083113}%
  \BibitemOpen
  \bibfield  {author} {\bibinfo {author} {\bibfnamefont {P.}~\bibnamefont
  {Fendley}},\ }\bibfield  {title} {\bibinfo {title} {Topological order from
  quantum loops and nets},\ }\href
  {https://doi.org/https://doi.org/10.1016/j.aop.2008.04.011} {\bibfield
  {journal} {\bibinfo  {journal} {Annals of Physics}\ }\textbf {\bibinfo
  {volume} {323}},\ \bibinfo {pages} {3113} (\bibinfo {year}
  {2008})}\BibitemShut {NoStop}%
\bibitem [{\citenamefont {Levin}\ and\ \citenamefont
  {Wen}(2005)}]{PhysRevB.71.045110}%
  \BibitemOpen
  \bibfield  {author} {\bibinfo {author} {\bibfnamefont {M.~A.}\ \bibnamefont
  {Levin}}\ and\ \bibinfo {author} {\bibfnamefont {X.-G.}\ \bibnamefont
  {Wen}},\ }\bibfield  {title} {\bibinfo {title} {String-net condensation: A
  physical mechanism for topological phases},\ }\href
  {https://doi.org/10.1103/PhysRevB.71.045110} {\bibfield  {journal} {\bibinfo
  {journal} {Phys. Rev. B}\ }\textbf {\bibinfo {volume} {71}},\ \bibinfo
  {pages} {045110} (\bibinfo {year} {2005})}\BibitemShut {NoStop}%
\bibitem [{\citenamefont {Baxter}(1985)}]{baxter2016exactly}%
  \BibitemOpen
  \bibfield  {author} {\bibinfo {author} {\bibfnamefont {R.~J.}\ \bibnamefont
  {Baxter}},\ }\href {https://doi.org/10.1142/9789814415255_0002} {\emph
  {\bibinfo {title} {Exactly Solved Models in Statistical Mechanics}}}\
  (\bibinfo  {publisher} {WORLD SCIENTIFIC},\ \bibinfo {year}
  {1985})\BibitemShut {NoStop}%
\bibitem [{\citenamefont {Dye}(2003)}]{Dye2003}%
  \BibitemOpen
  \bibfield  {author} {\bibinfo {author} {\bibfnamefont {H.~A.}\ \bibnamefont
  {Dye}},\ }\bibfield  {title} {\bibinfo {title} {Unitary solutions to the
  yang–baxter equation in dimension four},\ }\href
  {https://doi.org/10.1023/A:1025843426102} {\bibfield  {journal} {\bibinfo
  {journal} {Quantum Information Processing}\ }\textbf {\bibinfo {volume}
  {2}},\ \bibinfo {pages} {117} (\bibinfo {year} {2003})}\BibitemShut {NoStop}%
\bibitem [{\citenamefont {Ljubotina}\ \emph {et~al.}(2019)\citenamefont
  {Ljubotina}, \citenamefont {Zadnik},\ and\ \citenamefont
  {Prosen}}]{PhysRevLett.122.150605}%
  \BibitemOpen
  \bibfield  {author} {\bibinfo {author} {\bibfnamefont {M.}~\bibnamefont
  {Ljubotina}}, \bibinfo {author} {\bibfnamefont {L.}~\bibnamefont {Zadnik}},\
  and\ \bibinfo {author} {\bibfnamefont {T.}~\bibnamefont {Prosen}},\
  }\bibfield  {title} {\bibinfo {title} {Ballistic spin transport in a
  periodically driven integrable quantum system},\ }\href
  {https://doi.org/10.1103/PhysRevLett.122.150605} {\bibfield  {journal}
  {\bibinfo  {journal} {Phys. Rev. Lett.}\ }\textbf {\bibinfo {volume} {122}},\
  \bibinfo {pages} {150605} (\bibinfo {year} {2019})}\BibitemShut {NoStop}%
\bibitem [{\citenamefont {S\'a}\ \emph {et~al.}(2021)\citenamefont {S\'a},
  \citenamefont {Ribeiro},\ and\ \citenamefont {Prosen}}]{PhysRevB.103.115132}%
  \BibitemOpen
  \bibfield  {author} {\bibinfo {author} {\bibfnamefont {L.}~\bibnamefont
  {S\'a}}, \bibinfo {author} {\bibfnamefont {P.}~\bibnamefont {Ribeiro}},\ and\
  \bibinfo {author} {\bibfnamefont {T.}~\bibnamefont {Prosen}},\ }\bibfield
  {title} {\bibinfo {title} {Integrable nonunitary open quantum circuits},\
  }\href {https://doi.org/10.1103/PhysRevB.103.115132} {\bibfield  {journal}
  {\bibinfo  {journal} {Phys. Rev. B}\ }\textbf {\bibinfo {volume} {103}},\
  \bibinfo {pages} {115132} (\bibinfo {year} {2021})}\BibitemShut {NoStop}%
\bibitem [{\citenamefont {Nayak}\ \emph {et~al.}(2008)\citenamefont {Nayak},
  \citenamefont {Simon}, \citenamefont {Stern}, \citenamefont {Freedman},\ and\
  \citenamefont {Das~Sarma}}]{RevModPhys.80.1083}%
  \BibitemOpen
  \bibfield  {author} {\bibinfo {author} {\bibfnamefont {C.}~\bibnamefont
  {Nayak}}, \bibinfo {author} {\bibfnamefont {S.~H.}\ \bibnamefont {Simon}},
  \bibinfo {author} {\bibfnamefont {A.}~\bibnamefont {Stern}}, \bibinfo
  {author} {\bibfnamefont {M.}~\bibnamefont {Freedman}},\ and\ \bibinfo
  {author} {\bibfnamefont {S.}~\bibnamefont {Das~Sarma}},\ }\bibfield  {title}
  {\bibinfo {title} {Non-abelian anyons and topological quantum computation},\
  }\href {https://doi.org/10.1103/RevModPhys.80.1083} {\bibfield  {journal}
  {\bibinfo  {journal} {Rev. Mod. Phys.}\ }\textbf {\bibinfo {volume} {80}},\
  \bibinfo {pages} {1083} (\bibinfo {year} {2008})}\BibitemShut {NoStop}%
\bibitem [{\citenamefont {Bertini}\ \emph {et~al.}(2019)\citenamefont
  {Bertini}, \citenamefont {Kos},\ and\ \citenamefont
  {Prosen}}]{PhysRevLett.123.210601}%
  \BibitemOpen
  \bibfield  {author} {\bibinfo {author} {\bibfnamefont {B.}~\bibnamefont
  {Bertini}}, \bibinfo {author} {\bibfnamefont {P.}~\bibnamefont {Kos}},\ and\
  \bibinfo {author} {\bibfnamefont {T.}~\bibnamefont {Prosen}},\ }\bibfield
  {title} {\bibinfo {title} {Exact correlation functions for dual-unitary
  lattice models in $1+1$ dimensions},\ }\href
  {https://doi.org/10.1103/PhysRevLett.123.210601} {\bibfield  {journal}
  {\bibinfo  {journal} {Phys. Rev. Lett.}\ }\textbf {\bibinfo {volume} {123}},\
  \bibinfo {pages} {210601} (\bibinfo {year} {2019})}\BibitemShut {NoStop}%
\bibitem [{\citenamefont {Ippoliti}\ and\ \citenamefont
  {Khemani}(2021)}]{PhysRevLett.126.060501}%
  \BibitemOpen
  \bibfield  {author} {\bibinfo {author} {\bibfnamefont {M.}~\bibnamefont
  {Ippoliti}}\ and\ \bibinfo {author} {\bibfnamefont {V.}~\bibnamefont
  {Khemani}},\ }\bibfield  {title} {\bibinfo {title} {Postselection-free
  entanglement dynamics via spacetime duality},\ }\href
  {https://doi.org/10.1103/PhysRevLett.126.060501} {\bibfield  {journal}
  {\bibinfo  {journal} {Phys. Rev. Lett.}\ }\textbf {\bibinfo {volume} {126}},\
  \bibinfo {pages} {060501} (\bibinfo {year} {2021})}\BibitemShut {NoStop}%
\bibitem [{\citenamefont {Nakata}\ \emph {et~al.}(2021)\citenamefont {Nakata},
  \citenamefont {Takayanagi}, \citenamefont {Taki}, \citenamefont {Tamaoka},\
  and\ \citenamefont {Wei}}]{PhysRevD.103.026005}%
  \BibitemOpen
  \bibfield  {author} {\bibinfo {author} {\bibfnamefont {Y.}~\bibnamefont
  {Nakata}}, \bibinfo {author} {\bibfnamefont {T.}~\bibnamefont {Takayanagi}},
  \bibinfo {author} {\bibfnamefont {Y.}~\bibnamefont {Taki}}, \bibinfo {author}
  {\bibfnamefont {K.}~\bibnamefont {Tamaoka}},\ and\ \bibinfo {author}
  {\bibfnamefont {Z.}~\bibnamefont {Wei}},\ }\bibfield  {title} {\bibinfo
  {title} {New holographic generalization of entanglement entropy},\ }\href
  {https://doi.org/10.1103/PhysRevD.103.026005} {\bibfield  {journal} {\bibinfo
   {journal} {Phys. Rev. D}\ }\textbf {\bibinfo {volume} {103}},\ \bibinfo
  {pages} {026005} (\bibinfo {year} {2021})}\BibitemShut {NoStop}%
\bibitem [{\citenamefont {Hastings}\ and\ \citenamefont
  {Mahajan}(2015)}]{PhysRevA.91.032306}%
  \BibitemOpen
  \bibfield  {author} {\bibinfo {author} {\bibfnamefont {M.~B.}\ \bibnamefont
  {Hastings}}\ and\ \bibinfo {author} {\bibfnamefont {R.}~\bibnamefont
  {Mahajan}},\ }\bibfield  {title} {\bibinfo {title} {Connecting entanglement
  in time and space: Improving the folding algorithm},\ }\href
  {https://doi.org/10.1103/PhysRevA.91.032306} {\bibfield  {journal} {\bibinfo
  {journal} {Phys. Rev. A}\ }\textbf {\bibinfo {volume} {91}},\ \bibinfo
  {pages} {032306} (\bibinfo {year} {2015})}\BibitemShut {NoStop}%
\bibitem [{\citenamefont {Lerose}\ \emph {et~al.}(2021)\citenamefont {Lerose},
  \citenamefont {Sonner},\ and\ \citenamefont {Abanin}}]{PhysRevX.11.021040}%
  \BibitemOpen
  \bibfield  {author} {\bibinfo {author} {\bibfnamefont {A.}~\bibnamefont
  {Lerose}}, \bibinfo {author} {\bibfnamefont {M.}~\bibnamefont {Sonner}},\
  and\ \bibinfo {author} {\bibfnamefont {D.~A.}\ \bibnamefont {Abanin}},\
  }\bibfield  {title} {\bibinfo {title} {Influence matrix approach to many-body
  floquet dynamics},\ }\href {https://doi.org/10.1103/PhysRevX.11.021040}
  {\bibfield  {journal} {\bibinfo  {journal} {Phys. Rev. X}\ }\textbf {\bibinfo
  {volume} {11}},\ \bibinfo {pages} {021040} (\bibinfo {year}
  {2021})}\BibitemShut {NoStop}%
\bibitem [{\citenamefont {Foligno}\ \emph {et~al.}(2023)\citenamefont
  {Foligno}, \citenamefont {Zhou},\ and\ \citenamefont
  {Bertini}}]{PhysRevX.13.041008}%
  \BibitemOpen
  \bibfield  {author} {\bibinfo {author} {\bibfnamefont {A.}~\bibnamefont
  {Foligno}}, \bibinfo {author} {\bibfnamefont {T.}~\bibnamefont {Zhou}},\ and\
  \bibinfo {author} {\bibfnamefont {B.}~\bibnamefont {Bertini}},\ }\bibfield
  {title} {\bibinfo {title} {Temporal entanglement in chaotic quantum
  circuits},\ }\href {https://doi.org/10.1103/PhysRevX.13.041008} {\bibfield
  {journal} {\bibinfo  {journal} {Phys. Rev. X}\ }\textbf {\bibinfo {volume}
  {13}},\ \bibinfo {pages} {041008} (\bibinfo {year} {2023})}\BibitemShut
  {NoStop}%
\bibitem [{\citenamefont {Yao}\ and\ \citenamefont
  {Claeys}(2024)}]{PhysRevResearch.6.043077}%
  \BibitemOpen
  \bibfield  {author} {\bibinfo {author} {\bibfnamefont {J.}~\bibnamefont
  {Yao}}\ and\ \bibinfo {author} {\bibfnamefont {P.~W.}\ \bibnamefont
  {Claeys}},\ }\bibfield  {title} {\bibinfo {title} {Temporal entanglement
  barriers in dual-unitary clifford circuits with measurements},\ }\href
  {https://doi.org/10.1103/PhysRevResearch.6.043077} {\bibfield  {journal}
  {\bibinfo  {journal} {Phys. Rev. Res.}\ }\textbf {\bibinfo {volume} {6}},\
  \bibinfo {pages} {043077} (\bibinfo {year} {2024})}\BibitemShut {NoStop}%
\bibitem [{app()}]{appendix_note}%
  \BibitemOpen
  \href@noop {} {}\bibinfo {note} {See Supplemental Material for alternative
  definitions of the entanglement entropy, calculations for the Hopf link,
  algebra structure of gates, exact Renyi entropy and two-point correlation
  functions without measurement, additional data for the effect of random
  replacements by swap gates and computation methods for Clifford circuit and
  classical simulation used in the paper.}\BibitemShut {Stop}%
\bibitem [{\citenamefont {Kauffman}(2001)}]{10.1142/4256}%
  \BibitemOpen
  \bibfield  {author} {\bibinfo {author} {\bibfnamefont {L.~H.}\ \bibnamefont
  {Kauffman}},\ }\href {https://doi.org/10.1142/4256} {\emph {\bibinfo {title}
  {Knots and Physics}}},\ \bibinfo {edition} {3rd}\ ed.\ (\bibinfo  {publisher}
  {WORLD SCIENTIFIC},\ \bibinfo {year} {2001})\BibitemShut {NoStop}%
\bibitem [{\citenamefont {Gottesman}(1997)}]{gottesman1997stabilizer}%
  \BibitemOpen
  \bibfield  {author} {\bibinfo {author} {\bibfnamefont {D.}~\bibnamefont
  {Gottesman}},\ }\href@noop {} {\emph {\bibinfo {title} {Stabilizer codes and
  quantum error correction}}}\ (\bibinfo  {publisher} {California Institute of
  Technology},\ \bibinfo {year} {1997})\BibitemShut {NoStop}%
\bibitem [{\citenamefont {Aaronson}\ and\ \citenamefont
  {Gottesman}(2004)}]{PhysRevA.70.052328}%
  \BibitemOpen
  \bibfield  {author} {\bibinfo {author} {\bibfnamefont {S.}~\bibnamefont
  {Aaronson}}\ and\ \bibinfo {author} {\bibfnamefont {D.}~\bibnamefont
  {Gottesman}},\ }\bibfield  {title} {\bibinfo {title} {Improved simulation of
  stabilizer circuits},\ }\href {https://doi.org/10.1103/PhysRevA.70.052328}
  {\bibfield  {journal} {\bibinfo  {journal} {Phys. Rev. A}\ }\textbf {\bibinfo
  {volume} {70}},\ \bibinfo {pages} {052328} (\bibinfo {year}
  {2004})}\BibitemShut {NoStop}%
\bibitem [{\citenamefont {Nahum}\ \emph {et~al.}(2017)\citenamefont {Nahum},
  \citenamefont {Ruhman}, \citenamefont {Vijay},\ and\ \citenamefont
  {Haah}}]{PhysRevX.7.031016}%
  \BibitemOpen
  \bibfield  {author} {\bibinfo {author} {\bibfnamefont {A.}~\bibnamefont
  {Nahum}}, \bibinfo {author} {\bibfnamefont {J.}~\bibnamefont {Ruhman}},
  \bibinfo {author} {\bibfnamefont {S.}~\bibnamefont {Vijay}},\ and\ \bibinfo
  {author} {\bibfnamefont {J.}~\bibnamefont {Haah}},\ }\bibfield  {title}
  {\bibinfo {title} {Quantum entanglement growth under random unitary
  dynamics},\ }\href {https://doi.org/10.1103/PhysRevX.7.031016} {\bibfield
  {journal} {\bibinfo  {journal} {Phys. Rev. X}\ }\textbf {\bibinfo {volume}
  {7}},\ \bibinfo {pages} {031016} (\bibinfo {year} {2017})}\BibitemShut
  {NoStop}%
\bibitem [{\citenamefont {Nielsen}\ and\ \citenamefont
  {Chuang}(2010)}]{Nielsen_Chuang_2010}%
  \BibitemOpen
  \bibfield  {author} {\bibinfo {author} {\bibfnamefont {M.~A.}\ \bibnamefont
  {Nielsen}}\ and\ \bibinfo {author} {\bibfnamefont {I.~L.}\ \bibnamefont
  {Chuang}},\ }\href@noop {} {\emph {\bibinfo {title} {Quantum Computation and
  Quantum Information: 10th Anniversary Edition}}}\ (\bibinfo  {publisher}
  {Cambridge University Press},\ \bibinfo {year} {2010})\BibitemShut {NoStop}%
\bibitem [{\citenamefont {Sierant}\ and\ \citenamefont
  {Turkeshi}(2022)}]{sierant2022universal}%
  \BibitemOpen
  \bibfield  {author} {\bibinfo {author} {\bibfnamefont {P.}~\bibnamefont
  {Sierant}}\ and\ \bibinfo {author} {\bibfnamefont {X.}~\bibnamefont
  {Turkeshi}},\ }\bibfield  {title} {\bibinfo {title} {Universal behavior
  beyond multifractality of wave functions at measurement-induced phase
  transitions},\ }\href {https://doi.org/10.1103/PhysRevLett.128.130605}
  {\bibfield  {journal} {\bibinfo  {journal} {Phys. Rev. Lett.}\ }\textbf
  {\bibinfo {volume} {128}},\ \bibinfo {pages} {130605} (\bibinfo {year}
  {2022})}\BibitemShut {NoStop}%
\bibitem [{\citenamefont {Sun}\ and\ \citenamefont {Chen}(2025)}]{data}%
  \BibitemOpen
  \bibfield  {author} {\bibinfo {author} {\bibfnamefont {S.-K.}\ \bibnamefont
  {Sun}}\ and\ \bibinfo {author} {\bibfnamefont {S.}~\bibnamefont {Chen}},\
  }\bibfield  {title} {\bibinfo {title} {Code and data for "entanglement
  transition and suppression of critical phase of thermofield double state in
  monitored quantum circuit with unitary $r$ matrix gates"},\ }\href
  {https://github.com/Pknight-ssk/dataFor2503.00396} {\bibfield  {journal}
  {\bibinfo  {journal} {Github}\ } (\bibinfo {year} {2025})}\BibitemShut
  {NoStop}%
\end{thebibliography}%
	
\clearpage
\appendix
\onecolumngrid

\section{Alternative definitions of the entanglement entropy}
In this section, we provide some alternative definitions of the entanglement entropy used in the main text. We use $U$ to refer to the circuit with forced-measurements, which is a matrix product of unitary gates $R$, projectors $P$ and identity operators $I$ according to the brickwall pattern, $U = \prod_{t}\prod_{i=0,1}\otimes_{j}X^{t}_{2j-1+i, 2j+i}$ where $X^t$ is randomly chosen from $R$, $P$ and $I$ at time slice $t$, and the lower index indicates sites that $X^t$ acts on.
\begin{enumerate}
	\item Since projector $P$ affects the Frobenius norm of $U$, $\|U\|_2 := \sqrt{\tr \lp U^{\dagger}U \rp}$, we define the normalized circuit $U'$ as $U'=U/\|U\|_2$. Then, by singular value decomposition (SVD)  $U' = M \Gamma N^{\dagger}$, we define the entanglement of a single trajectory as $S = -\tr \lp \Gamma^2 \log_2 \Gamma^2 \rp$.
	\item  Since the final state after the application of the circuit is a pure state by renormalization, we write the final state formally as $U\otimes I \ket{\Psi} =\sum_{i} c_i \ket{\psi_i}_1 \otimes \ket{\phi_i}_2$ (Schmidt decomposition), and we have
	\begin{equation}\label{eq:rho_form}
		U\otimes I \, \ket{\Psi}\bra{\Psi} \, U^{\dagger}\otimes I = \sum_{i,j} c_i c_j^{*} \ket{\psi_i} \bra{\psi_j}_1 \otimes \ket{\phi_i}\bra{\phi_j}_2 \;\; .
	\end{equation}
	Utilizing Choi representation of a channel,
	\begin{equation}
		J_U (\Phi) \equiv U\otimes I \, \ket{\Psi}\bra{\Psi} \, U^{\dagger}\otimes I = \sum_{i,j} \Phi\lp \ket{i} \bra{j}_1\rp \otimes \ket{i} \bra{j}_2 \;\; ,
	\end{equation}
	where $\Phi$ represents the channel of the circuit. The von Neumann entropy $S=-\sum_i c_i^2 \log_2(c_i^2)$ of the time-evolved TFDS $U\otimes I \ket{\Psi} =\sum_{i} c_i \ket{\psi_i}_1 \otimes \ket{\phi_i}_2$ in the main text corresponds to 
	\begin{equation}
		\begin{aligned}
			\rho(\Phi)_1 &\equiv \textbf{Tr}_2 \lp J_U(\Phi) \rp = \sum_{i} c_i^2 \ket{\psi_i} \bra{\psi_i}_1 \\
			S\lp J_U (\Phi) \rp & = - \textbf{Tr} \lp \rho(\Phi)_1 \log_2 (\rho(\Phi)_1) \rp \;\; .
		\end{aligned}
	\end{equation}
	This is a reasonable definition, because $\rho(\Phi)_1$ is a positive-semidefinitive hermitian matrix, thus a density matrix.
\end{enumerate}

\section{Computation of topological invariant by R matrix}
The invariant of Hopf link is computed as
\begin{equation}
	\begin{aligned}
		\tau \lp \begin{tikzpicture}[baseline=(current  bounding  box.center), scale=0.2]
			\draw[thick] (-1,-1) -- (1,1);
			\draw[thick] (-1, 1) -- (-0.2,0.2);
			\draw[thick](0.2, -0.2) -- (1,-1);
			
			\draw[thick] (3,-1) -- (5,1);
			\draw[thick] (3, 1) -- (3.8,0.2);
			\draw[thick](4.2, -0.2) -- (5,-1);
			
			\draw[thick] (3,1) arc [start angle= 45, end angle=135, radius=1.414];
			\draw[thick] (3,-1) arc [start angle= -45, end angle=-135, radius=1.414];
			
			\draw[thick] (3,-2.5) arc [start angle= 45, end angle=135, radius=1.414];
			\draw[thick] (3,2.5) arc [start angle= -45, end angle=-135, radius=1.414];
			
			\draw[thick] (-1,2.5) arc [start angle= 135, end angle=45, radius=1.414];
			\draw[thick] (3,2.5) arc [start angle= 135, end angle=45, radius=1.414];
			\draw[thick] (-1,-2.5) arc [start angle= -135, end angle=-45, radius=1.414];
			\draw[thick] (3,-2.5) arc [start angle= -135, end angle=-45, radius=1.414];
			
			\draw[thick] (-1,1) -- (-1,2.5);
			\draw[thick] (5,1) -- (5,2.5);
			
			\draw[thick] (-1,-1) -- (-1,-2.5);
			\draw[thick] (5,-1) -- (5,-2.5);
			
		\end{tikzpicture} \rp
		&= 
		\tau \lp \begin{tikzpicture}[baseline=(current  bounding  box.center), scale=0.2]
			\draw[thick] (-1,-1) -- (1,1);
			\draw[thick] (-1, 1) -- (-0.2,0.2);
			\draw[thick](0.2, -0.2) -- (1,-1);
			
			\draw[thick] (3,-1) -- (3.8,-0.2);
			\draw[thick] (3, 1) -- (5,-1);
			\draw[thick](4.2, 0.2) -- (5,1);
			
			\draw[thick] (3,1) arc [start angle= 45, end angle=135, radius=1.414];
			\draw[thick] (3,-1) arc [start angle= -45, end angle=-135, radius=1.414];
			
			\draw[thick] (3,-2.5) arc [start angle= 45, end angle=135, radius=1.414];
			\draw[thick] (3,2.5) arc [start angle= -45, end angle=-135, radius=1.414];
			
			\draw[thick] (-1,2.5) arc [start angle= 135, end angle=45, radius=1.414];
			\draw[thick] (3,2.5) arc [start angle= 135, end angle=45, radius=1.414];
			\draw[thick] (-1,-2.5) arc [start angle= -135, end angle=-45, radius=1.414];
			\draw[thick] (3,-2.5) arc [start angle= -135, end angle=-45, radius=1.414];
			
			\draw[thick] (-1,1) -- (-1,2.5);
			\draw[thick] (5,1) -- (5,2.5);
			
			\draw[thick] (-1,-1) -- (-1,-2.5);
			\draw[thick] (5,-1) -- (5,-2.5);
			
		\end{tikzpicture} \rp \; + \frac{2i}{\sqrt{1+c^2}} \;
		\tau \lp \begin{tikzpicture}[baseline=(current  bounding  box.center), scale=0.2]
			\draw[thick] (-1,-1) -- (1,1);
			\draw[thick] (-1, 1) -- (-0.2,0.2);
			\draw[thick](0.2, -0.2) -- (1,-1);
			
			\draw[thick] (3,1) arc [start angle= 45, end angle=-45, radius=1.414];
			\draw[thick] (5,1) arc [start angle= 135, end angle=225, radius=1.414];
			
			\draw[thick] (3,1) arc [start angle= 45, end angle=135, radius=1.414];
			\draw[thick] (3,-1) arc [start angle= -45, end angle=-135, radius=1.414];
			
			\draw[thick] (3,-2.5) arc [start angle= 45, end angle=135, radius=1.414];
			\draw[thick] (3,2.5) arc [start angle= -45, end angle=-135, radius=1.414];
			
			\draw[thick] (-1,2.5) arc [start angle= 135, end angle=45, radius=1.414];
			\draw[thick] (3,2.5) arc [start angle= 135, end angle=45, radius=1.414];
			\draw[thick] (-1,-2.5) arc [start angle= -135, end angle=-45, radius=1.414];
			\draw[thick] (3,-2.5) arc [start angle= -135, end angle=-45, radius=1.414];
			
			\draw[thick] (-1,1) -- (-1,2.5);
			\draw[thick] (5,1) -- (5,2.5);
			
			\draw[thick] (-1,-1) -- (-1,-2.5);
			\draw[thick] (5,-1) -- (5,-2.5);
			
		\end{tikzpicture}\rp \; - \frac{2i}{\sqrt{1+c^2}} \;
		\tau \lp \begin{tikzpicture}[baseline=(current  bounding  box.center), scale=0.2]
			\draw[thick] (-1,-1) -- (1,1);
			\draw[thick] (-1, 1) -- (-0.2,0.2);
			\draw[thick](0.2, -0.2) -- (1,-1);
			
			\draw[thick] (3,1) arc [start angle= -135, end angle=-45, radius=1.414];
			\draw[thick] (5,-1) arc [start angle= 45, end angle=135, radius=1.414];
			
			\draw[thick] (3,1) arc [start angle= 45, end angle=135, radius=1.414];
			\draw[thick] (3,-1) arc [start angle= -45, end angle=-135, radius=1.414];
			
			\draw[thick] (3,-2.5) arc [start angle= 45, end angle=135, radius=1.414];
			\draw[thick] (3,2.5) arc [start angle= -45, end angle=-135, radius=1.414];
			
			\draw[thick] (-1,2.5) arc [start angle= 135, end angle=45, radius=1.414];
			\draw[thick] (3,2.5) arc [start angle= 135, end angle=45, radius=1.414];
			\draw[thick] (-1,-2.5) arc [start angle= -135, end angle=-45, radius=1.414];
			\draw[thick] (3,-2.5) arc [start angle= -135, end angle=-45, radius=1.414];
			
			\draw[thick] (-1,1) -- (-1,2.5);
			\draw[thick] (5,1) -- (5,2.5);
			
			\draw[thick] (-1,-1) -- (-1,-2.5);
			\draw[thick] (5,-1) -- (5,-2.5);
			
		\end{tikzpicture} \rp \\
		&= n^2+ \frac{2i}{\sqrt{1+c^2}} k_{-} n -  \frac{2i}{\sqrt{1+c^2}} k_{+} n \\
		&=  4+ 2\frac{2i}{\sqrt{1+c^2}} \frac{2ic}{\sqrt{1+c^2}} \\
		&=  \frac{4(c-1)^2}{1+c^2} 
	\end{aligned}
\end{equation}
In first equality we wrote a $R$ as a sum of $R^{\dagger}$, $I$ and $P'$. In second equality we used delooping relations in (\ref{eq:BMW}). This result shows our invariant is somewhat different from other invariant polynomials. It is proportional to the inner product $\braket{\psi|P_{23} R_{12} R_{34} P_{23}|\psi}$, where $\ket{\psi}$ is $\frac{1}{2}\ket{\uparrow\uparrow+\downarrow\downarrow}_{12} \ket{\uparrow\uparrow+\downarrow\downarrow}_{34}$, and the factor is $n^4=16$.

\section{Algebraic relations of R matrix}
With the decomposition 
\begin{equation}
	\begin{aligned}
		R &= \frac{1}{\sqrt{1+c^2}} \lp i* I - i * P' + c * \text{SWAP}  \rp \\
		R^{\dagger} &= \frac{1}{\sqrt{1+c^2}} \lp -i*I+i *  P' + c * \text{SWAP}   \rp \;\;,
	\end{aligned}
\end{equation}
and matrix representation of $R$ where $\phi = \text{arccot}(c)$ and $V$ is an unitary rotation along $X$ direction, $V=e^{i \frac{\pi}{4}X_1}e^{i \frac{\pi}{4}X_2}$, by which the $R(c)$ gate reduces to $e^{-\frac{i \pi}{4}H}$ where $H$ is generally a two-site $XXZ$ hamiltonian, $H=X_1X_2+Y_1Y_2-\lp \frac{4}{\pi}\text{arccot} (c) -1 \rp Z_1Z_2-I$,
\begin{equation}
	R(c)=\frac{1}{\sqrt{1+c^2}} \begin{pmatrix}
		c & 0 & 0 & -i \\
		0 & i & c & 0 \\
		0 & c & i & 0 \\
		-i & 0 & 0 & c
	\end{pmatrix} \xrightarrow{V} \begin{pmatrix}
		e^{i \phi} & 0 & 0 & 0 \\
		0 & 0 & e^{-i \phi} & 0 \\
		0 & e^{-i \phi} & 0 & 0 \\
		0 & 0 & 0 & e^{i \phi}
	\end{pmatrix}
\end{equation}
as well as diagrammatic representations of $R$, $P'$, $I$ 
and $S$ (swap gate)
\begin{equation}\label{eq:graphical P and I}
	\begin{aligned}
		P' &:= \raisebox{0.2em}{\begin{tikzpicture}[baseline=(current  bounding  box.center), scale=0.8]
				\draw[thick] (-1,-1) arc [start angle= -135, end angle=-45, radius=0.707];
				\draw[thick] (-1,-2) arc [start angle= 135, end angle=45, radius=0.707];
		\end{tikzpicture}} = \ket{\uparrow\uparrow+\downarrow\downarrow}\bra{\uparrow\uparrow+\downarrow\downarrow} = \begin{pmatrix}
			1 & 0 & 0 & 1 \\
			0 & 0 & 0 & 0 \\
			0 &0& 0 & 0 \\
			1 & 0 & 0 & 1
		\end{pmatrix}  \\
		I &:= \raisebox{0.2em}{\begin{tikzpicture}[baseline=(current  bounding  box.center), scale=0.8]
				\draw[thick] (0,0) arc [start angle= -45, end angle=45, radius=0.707];
				\draw[thick] (1,0) arc [start angle= -135, end angle=-225, radius=0.707];
		\end{tikzpicture}} = \begin{pmatrix}
			1 & 0 & 0 & 0 \\
			0 & 1 & 0 & 0 \\
			0 & 0 &1 & 0 \\
			0 & 0 & 0 & 1
		\end{pmatrix}\\
		\text{SWAP} &:= \raisebox{0.2em}{\begin{tikzpicture}[baseline=(current  bounding  box.center), scale=0.4]
				\draw[thick] (-1,-1) -- (1,1);
				\draw[thick] (-1, 1) -- (1,-1);
				\draw[thick, fill=myred] (0,0) circle [radius=0.2];
		\end{tikzpicture}} = \begin{pmatrix}
			1 & 0 & 0 & 0 \\
			0 & 0 & 1 & 0 \\
			0 &1 & 0 & 0 \\
			0 & 0 & 0 & 1
		\end{pmatrix} \;\; ,
	\end{aligned}
\end{equation}
it is straight forward to verify that they form an algebra. We denote $X_i$ for the $X$ element that acts on site $i$ and $i+1$. First, $P'$ alone is a Temperley-Lieb generator and if combined with $S$, which is a generator of symmetric group, then they are generators of a Brauer algebra as they satisfy
\begin{equation}\label{eq:Brauer algebra}
	\begin{aligned}
		&S_i^2=I_i , \; {P'_i}^2=2{P'_i}\\
		&[S_i, S_j] = [S_i, {P'_j}] = [{P'_i}, {P'_j}] = 0   \; \text{whenever} \; |i-j|>1\\
		&S_i S_{i+1} S_i = S_{i+1}S_i S_{i+1}, \; 
		{P'_i}{P'_{i+1}}{P'_i}={P'_i}\\
		&S_i S_{i\pm1}{P'_i}={P'_{i\pm1}}{P'_i}, \;
		{P'_i}S_{i\pm1}S_i = {P'_i}{P'_{i\pm1}}\\
		&S_i {P'_i} = {P'_i} S_i = {P'_i},\;
		{P'_i}S_{i\pm1}{P'_i}={P'_i} \;.
	\end{aligned}
\end{equation}
Since swap gate is a special version of $R$ ($|c|\rightarrow \infty$), it is natural to consider the algebra generated by $R$, $P'$ and $I$. They generates the Birman–Murakami–Wenzl (BMW) algebra, as they satisfy (\ref{eq:Brauer algebra}) except that
\begin{equation}\label{eq:BMW}
	\begin{aligned}
		&R_i^2\neq I_i \\
		& R_j-R_j^{\dagger}=\frac{2i}{\sqrt{1+c^2}}\lp I_j - {P'_j} \rp \;\; \text{(Skein relation) }\\
		&R_i {P'_i} = {P'_i} R_i = k_{+} {P'_i} \;\; \text{(Delooping relations) }\\
		&{P'_i}R_{i\pm1}{P'_i}=k_{-}{P'_i} \;\; \text{(Delooping relations) }\\
		& R_{\pm1}{P'_i}R_{\pm1}=R_i^{\dagger}{P'_{\pm1}}R_i^{\dagger}\\
		&R_{\pm1}{P'_i}{P'_{\pm1}}=R_i^{\dagger} {P'_{\pm1}} , \; {P'_{\pm1}}{P'_i}R_{\pm1}={P'_{\pm1}}R_i^{\dagger} 
	\end{aligned}
\end{equation}
where $k_{+} = e^{-i \theta}$, $k_{-} = e^{i \theta}$, $\theta = \arctan (c)$. This version of BMW algebra is not the original definition but modified to agrees with Kauffman's link invariant, and then it is isomorphic to Kauffman's tangle algebra. All the above can be easily checked using a diagrammatic representation or matrix multiplication. The circuit then is just an element of this algebra as it only involves the product of $R$, $P'$ and $I$.

\section{Solvable correlation and exact Rényi entropy without measurements}
\begin{figure}[tb]
	\centering
	\includegraphics[width=0.45\linewidth]{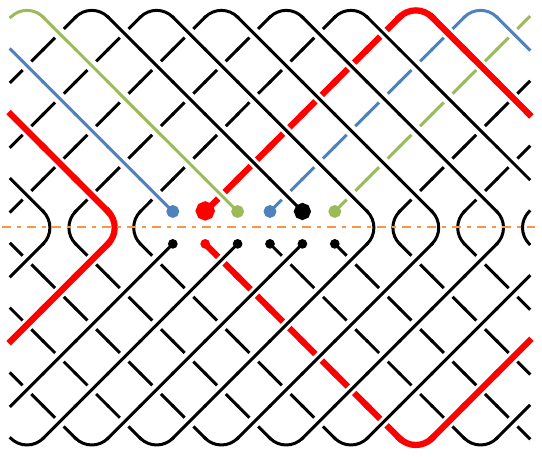}
	\caption{(Color online) An example for the calculation of the second order Rényi entropy. The system length $L=16$, the subsystem length $L_A = 6$ and the system evolves for 3 time steps. The 2 sites that connect to $\bar{A}$ are marked by larger dots and one of their worldlines is marked by a thickened red line. The horizontal dashed line separates the forward and backward evolution $U$ and $U^{\dagger}$. }
	\label{fig:figrenyi2}
\end{figure}
Since our $R$ matrix has graphical representation as a braid group generator, it is clearly dual-unitary. The advantage of dual-unitary gates is we can exactly compute the correlation functions. Moreover if we consider periodic boundary condition and the initial states are either infinite temperature state or tensor product of Bell states $\otimes \lp \ket{\uparrow}_i \ket{\uparrow}_{i^\prime}+\ket{\downarrow}_i \ket{\downarrow}_{i^\prime} \rp$, then the exact Rényi entropy $S_n(t) = \frac{1}{1-n} \log_2 \text{Tr} \lp \rho_A^n (t) \rp$ of subsystem $A$ is also known. For simplicity let us focus on the second order Rényi entropy. It equals to the number of sites that connect to the complement part $\bar{A}$. A simple example is shown in Fig.~\ref{fig:figrenyi2}, where the initial state is the tensor product of 8 nearest-neighbor Bell states and the evolution time is 3. Since there are only 2 sites connecting to $\bar{A}$, $S_2=2$.

To see this, just note that there is no link between any two loops, which means we can separate them without cut. Thus $\text{Tr} \lp \rho_A^2 (t) \rp = 2^{\#(\text{loops}) - L}$, only depends on the number of loops formed by concatenating two replicas. If a site in $A$ connects to $\bar{A}$, it will only contribute to one loop. So the overall Rényi entropy equals to the number of sites that connect to the complement part $\bar{A}$. Of course, if at some site $R$ randomly flipped to $R^{\dagger}$, there will be links and the above number of sites gives an upper bound for Rényi entropy.

\section{Additional data around phase boundaries}
\begin{figure}[tb]
	\centering
	\includegraphics[width=0.6\linewidth]{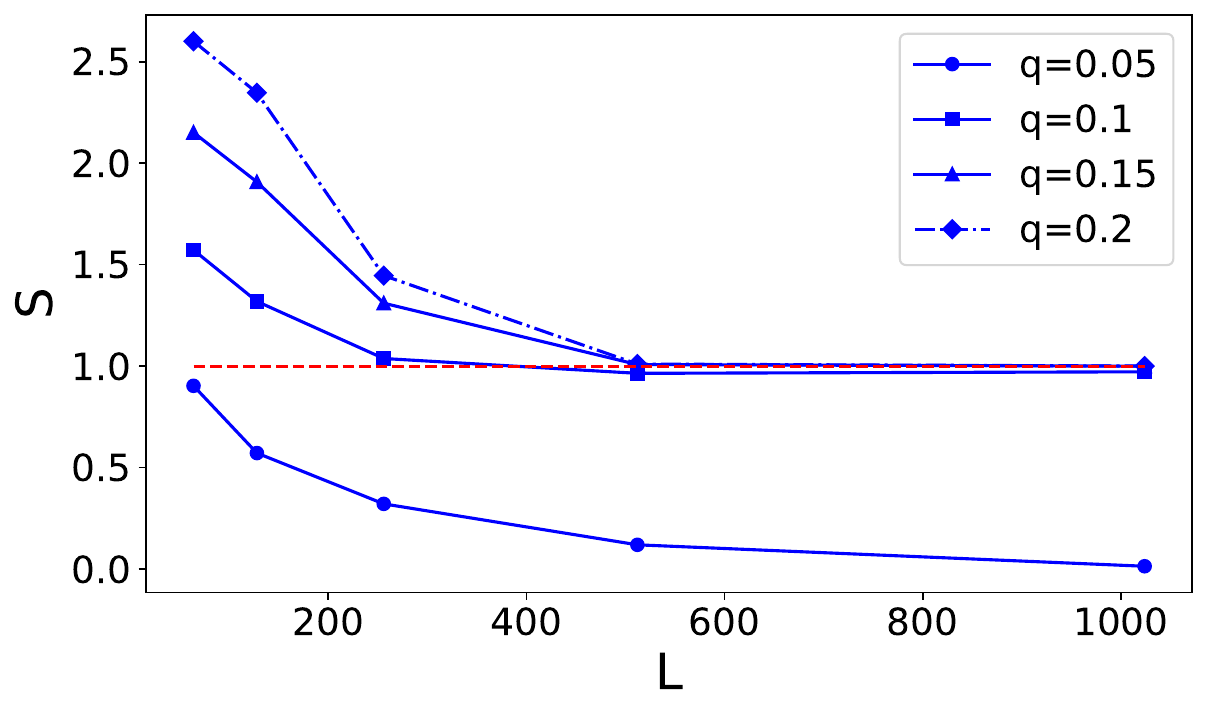}
	\caption{Some data for points near phase boundary of CPLC at $p=0.5$, $r=0.1$ and $t=L$. When $q=0.2$ the point is in critical phase and the other three are in area-law phase, but $q=0.05$ shows a different behavior as its entanglement approaches 0, since it is more far away from the transition boundary, compared to $q=0.1,0.15$. Red dashed line is a reference line for $S=1$. All data points are averaged from 16384 samples.}
	\label{fig:entp0d5-newregion}
\end{figure}

In this appendix we provide some additional data around phase boundaries, which shows evidence for suspected enriched phase or modified phase boundary at non-Clifford points. We set $r=0.1$. As shown in Fig.~\ref{fig:entp0d5-newregion}, for data points near phase boundary of CPLC at $p=0.5$ (the transition point is around 0.175), they converge to $S=1$ (except for $q=0.05$), but for $q=0.2$, which is in critical phase, its entanglement is strictly larger than 1. The converged value of entanglement of $q=0.1, 0.15$ is slightly less than 1. In contrast, when $q=0.05$, or $p=0.2$, $q=0.2$ (not shown), the converged value is 0. We conjecture this is evidence for enriched phases or modified phase boundary at non-Clifford points.

\section{Computation method}\label{sec:computation}
For numerical simulation, we adopt a random sampling of trajectories and we average over all trajectories to get entanglement entropy. The probability for each gate to be either a $R$ or $P$ is independent from each other. This independence makes the direct simulation of the model feasible. In the following we introduce two methods we used for previous results.

\subsection{Knitting and shuffling method for classical simulation}
We use this method to simulate the circuit at Clifford points. In this case since each trajectory is only a pairing configuration plus worldline length distribution, it can be efficiently computed by updating a tableau similar to the stabilizer formalism. There are three kinds of continuous worldline with ends: 1. both ends are on final time boundary; 2. both ends are on initial time boundary; 3. two ends are on different boundary. The first two cases are pairings of sites and the third case contributes to the spanning number, which counts how many worldlines stretch through different time boundary and upper bounds the entanglement entropy of the TFDS. A pairing configuration of the state is a list of integer. The number $a_i$ at site $i$ stands for the pairings $(i, a_i)$, that is site $i$ connects to site $a_i$. For example, (2,1,4,3) stands for the pairings (1,2) and (3,4). The length of worldline that starts from site $i$ is recorded in a different list at site $i$. It is efficient to first prepare some stripes of size $(L, 2)$ and then recursively apply a knitting and shuffling method, which will reduce the total complexity to $L\ln L$, instead of $L^2$. This reduction works because the only important information is how boundaries are connected and length distribution of worldlines, which we can easily read by moving a finger along worldlines. Every time we concatenate randomly two stripes to get a size-doubled stripe and record loops formed by this process. This operation takes $O(L)$ time. The total number of operation is $\ln L$, thus leading to a $O(L\ln L)$ time and space complexity (the number of loops is $O(L\ln L)$ because every concatenation produces at most $L/2$ loops).

\subsection{Stabilizer method for Clifford cases}
The circuit is Clifford when $c=0, \pm 1$ and $|c|\rightarrow \infty$. In these cases we can use stabilizer to simulate the time evolution efficiently on a classical computer. The stabilizer method is $O(L^2)$ if we discard measurement results, so we cannot reach a system size as large as in knitting and shuffling method. But this method provides us a direct approach to exact entanglement. For projective measurements, we first project to $XX$, and project to $ZZ$ forthwith. For $R(c=1)$ gate, it is Clifford because it maps $X_1\rightarrow-Y_1 Z_2$, $X_2\rightarrow -Z_1Y_2$, $Z_1\rightarrow Y_1 X_2$, $Z_2\rightarrow X_1 Y_2$.

\end{document}